\documentclass[numberedappendix]{emulateapj}
\usepackage{amssymb,amsmath}
\usepackage{verbatim}
\usepackage{color,hyperref}
\usepackage{needspace}
\definecolor{linkcolor}{rgb}{0,0,0.25}
\hypersetup{
  colorlinks=true,        
  linkcolor=linkcolor,    
  citecolor=linkcolor,    
  filecolor=linkcolor,    
  urlcolor=linkcolor      
}
\newcounter{address}
\newcommand{\ie}{i.e.}
\newcommand{\etal}{et al.}

\newcommand{\eg}{e.g.}

\renewcommand{\tablename}{Table}
\renewcommand{\figurename}{Figure}

\newcommand{\sectionname}{$\mathsection$}

\newcommand{\feh}{\ensuremath{[\mathrm{Fe/H}]}}
\newcommand{\afe}{\ensuremath{[\alpha\mathrm{/Fe}]}}

\newcommand{\dex}{\ensuremath{\,\mathrm{dex}}}

\newcommand{\Myr}{\ensuremath{\,\mathrm{Myr}}}
\newcommand{\Gyr}{\ensuremath{\,\mathrm{Gyr}}}

\newcommand{\kms}{\ensuremath{\,\mathrm{km\ s}^{-1}}}
\newcommand{\msun}{\ensuremath{\,\mathrm{M}_{\odot}}}

\newcommand{\apogee}{APOGEE}

\newcommand{\logg}{\ensuremath{\log g}}
\newcommand{\teff}{\ensuremath{T_{\mathrm{eff}}}}
\newcommand{\teffi}{\ensuremath{T_{\mathrm{eff},i}}}
\newcommand{\xh}{\ensuremath{[\mathrm{X/H}]}}
\newcommand{\xfe}{\ensuremath{[\mathrm{X/Fe}]}}

\submitted{}

\begin{document}

\title{The chemical homogeneity of open clusters}

\author{Jo~Bovy}
\affil{Department of Astronomy and Astrophysics, University of Toronto, 50
  St.  George Street, Toronto, ON, M5S 3H4, Canada;
  bovy@astro.utoronto.ca~}

\begin{abstract}
  Determining the level of chemical homogeneity in open clusters is of
  fundamental importance in the study of the evolution of star-forming
  clouds and that of the Galactic disk. Yet limiting the initial
  abundance spread in clusters has been hampered by difficulties in
  obtaining consistent spectroscopic abundances for different stellar
  types. Without reference to any specific model of stellar
  photospheres, a model for a homogeneous cluster is that it forms a
  one-dimensional sequence, with any differences between members due
  to variations in stellar mass and observational uncertainties. I
  present a novel method for investigating the abundance spread in
  open clusters that tests this one-dimensional hypothesis at the
  level of observed stellar spectra, rather than constraining
  homogeneity using derived abundances as traditionally done. Using
  high-resolution APOGEE spectra for 49 giants in M67, NGC 6819, and
  NGC 2420 I demonstrate that these spectra form one-dimensional
  sequences for each cluster. With detailed forward modeling of the
  spectra and Approximate Bayesian Computation, I derive strong limits
  on the initial abundance spread of 15 elements: $<0.01\,(0.02)\dex$
  for C and Fe, $\lesssim0.015\,(0.03)\dex$ for N, O, Mg, Si, and Ni,
  $\lesssim0.02\,(0.03)\dex$ for Al, Ca, and Mn, and
  $\lesssim0.03\,(0.05)\dex$ for Na, S, K, Ti, and V (at 68\,\% and
  95\,\% confidence, respectively). The strong limits on C and O imply
  that no pollution by massive core-collapse supernovae occurred
  during star formation in open clusters, which, thus, need to form
  within $\lesssim6\Myr$. Further development of this and related
  techniques will bring the power of differential abundances to stars
  other than solar twins in large spectroscopic surveys and will help
  unravel the history of star formation and chemical enrichment in the
  Milky Way through chemical tagging.
\end{abstract}


\section{Introduction}

The surface abundances of long-lived stars observed through
high-resolution spectroscopy hold the archaeological record of the
conditions of their formation. Carefully uncovering this history
through analyses of the observed spectroscopic, photometric, and
astrometric data has the potential to lead to transformative insights
into the nature of star formation, the evolution of massive stars, and
the detailed chemical and dynamical evolution of galactic disks. Yet,
after many decades of work on the theory of stellar photospheres and
orders of magnitude improvements in the quantity, quality, and variety
of observed stellar spectra, stellar spectroscopy remains challenging
due to incomplete theoretical models and the difficulty of taking into
account the many instrumental factors affecting observations. Because
of this, abundance uncertainties are still routinely quoted as being
``0.1\dex'', a seemingly magic number even though in practice
observational setups vary widely.

Measuring stellar abundances for many different elements with
uncertainties $\ll0.1\dex$ opens up a wide range of questions to
scientific investigation. Stars are believed to form in groups in
molecular clouds \citep[\eg,][]{Shu87a,Lada03a}, but exactly how the
intracloud medium evolves and mixes and how star formation proceeds in
such clouds, especially on timescales of a few Myr
\citep{McKee02a,Feng14a}, is difficult to study observationally
because the young clusters are mostly obscured from view. Determining
the spread (or tight limits $\ll0.1\dex$ on it) in the abundances of
elements produced on short timescales by Type II supernovae in
surviving clusters would provide strong constraints on analytic and
numerical work in this area.

Beyond the individual star clusters, stellar abundances of long-lived
stars trace the history of star formation, chemical enrichment, and
the interstellar medium. If the majority of stars are born in clusters
with tens of thousands of members sharing the same initial abundances,
we might be able to chemically tag individual star-formation events in
the Milky Way by determining abundances for large samples of stars
\citep{Freeman02a}. If successful, this tagging would provide the
chemical and dynamical history of the Milky Way's disk at a level of
detail far surpassing our currently limited, broad-brush picture
\citep[\eg,][]{BlandHawthorn10a}. To determine whether chemical
tagging is possible, three essential questions remain to be answered:
(a) What is the level of initial abundance spread in star clusters?
(b) Can we measure the variations between the chemical signatures of
different clusters to the level determined in (a) in light of
observational uncertainties and the effects of stellar evolution on
the present-day surface abundances? And (c) do different star-forming
clusters have chemical signatures that are sufficiently unique to
distinguish each star-forming event, given the ``chemical resolution''
attained in (b)?

In this paper I present a novel method for addressing question (a)
above through observations of the abundance spread in open clusters
and use it to determine the most stringent constraints on the chemical
homogeneity of open clusters to date. Old open clusters (with ages
$>1\Gyr$) are those rare star-formation remnants that have not been
destroyed yet by encounters with molecular clouds. As such, they may
constitute a biased sample of the full initial open-cluster
population. But it is also likely that an unbiased subset survives the
presumably random interactions with gravitational inhomogeneities, in
which case they can shed light on the properties of all of the
clusters that stars form in.

Previous work has established that open clusters are homogeneous at
the level of $\approx0.05$ to $0.1\dex$
\citep[\eg,][]{DeSilva06a,DeSilva07a,DeSilva07b,Reddy12a,Ting12a},
although these analyses typically proceed by comparing the observed
scatter to the estimated uncertainties, rather than inferring rigorous
limits on the scatter. The advent of large surveys of open clusters
has allowed such analyses to be performed for many clusters and many
different atomic species, with limits on the dispersion now routinely
reaching $\lesssim0.05\dex$ \citep{Blanco15a}. One of the main
limiting factors in these studies is the inability to measure
abundances on a consistent scale for different stellar types (\eg,
dwarfs, sub-giants, giants), reducing the number of stars available
for any analysis.

\subsection{Limiting intrinsic scatter}

\begin{figure}
  \includegraphics[width=0.48\textwidth,clip=]{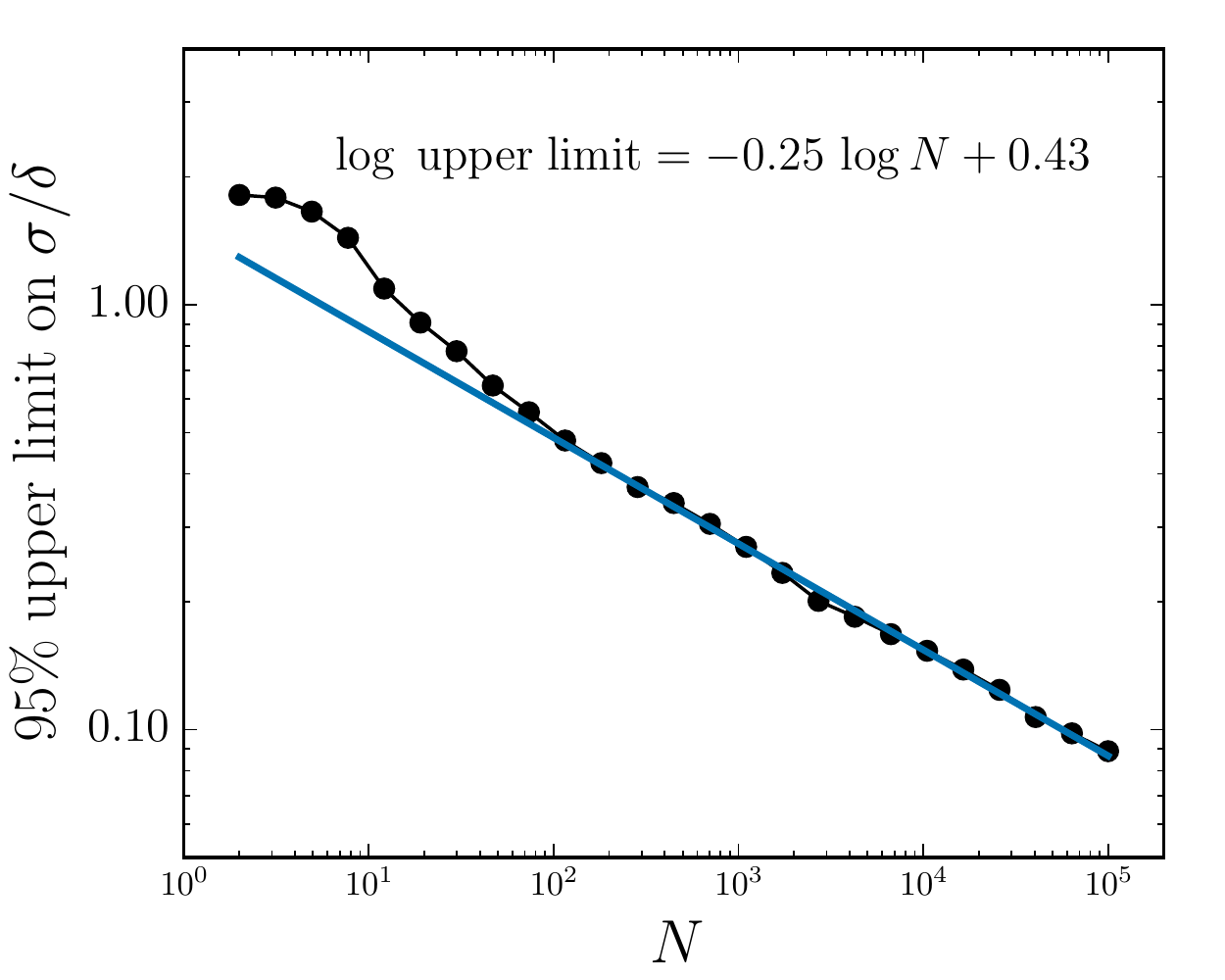}
\caption{The 95\,\% upper limit on the intrinsic scatter $\sigma$ for
  data drawn with uncertainty $\delta$ as a function of the number of
  data points $N$. Each point corresponds to the median of 50
  different mock data sets. The blue line is a fit to the large-$N$
  points, with the best-fit parameters indicated in the top right,
  demonstrating the asymptotic $N^{-1/4}$ dependence of the upper
  limit on $\sigma/\delta$ derived analytically in the text. For
  $0.10\dex$ abundance uncertainties, approximately 100,000 data
  points would be necessary to limit an open cluster's intrinsic
  scatter to below $0.01\dex$.}\label{fig:gaussdisp}
\end{figure}

Most work on determining or limiting the abundance spread in open
clusters does not carefully track the observational uncertainties,
even though these are key to establishing whether any measured scatter
is real or simply due to measurement errors. To illustrate this,
consider a simple experiment where $N$ mock data points $d_i$ with no
intrinsic scatter are drawn with Gaussian uncertainties with variance
$\delta^2$. The likelihood for the intrinsic scatter $\sigma$ when it
is assumed to be Gaussian is
\begin{equation}
\mathcal{L}(\sigma) \propto (\delta^2+\sigma^2)^{-N/2}\,\exp\left(-\frac{1}{2}\,\frac{\sum_{i=1}^N d_i^2}{\delta^2+\sigma^2}\right)\,.
\end{equation}
This is a $\chi^2$ distribution for the parameter $Q =
\frac{\sum_{i=1}^N d_i^2}{\delta^2+\sigma^2}$ with $N+2$ degrees of
freedom. In the large-$N$ limit, this distribution approaches a
Gaussian distribution with
\begin{equation}
  x = (Q-N-2)/\sqrt{2(N+2)} \sim \mathcal{N}(0,1)\,,
\end{equation}
where $\mathcal{N}(0,1)$ is the unit normal distribution. An upper
limit $\sigma_{\mathrm{ul}}$ on $\sigma$ at some confidence level
corresponds to a lower limit $Q_{\mathrm{ll}}$ on $Q$; in this case $x$ is
equal to some constant $-C$, \ie,
\begin{equation}
  (Q_{\mathrm{ll}}-N-2)/\sqrt{2(N+2)} = -C\,.
\end{equation}
For large $N$, $\sum_{i=1}^N d_i^2\approx \delta^2\,N$ and assuming
that $\sigma^2 \ll \delta^2$ we find for the upper limit
$\sigma_{\mathrm{ul}}$ on $\sigma$
\begin{equation}
  \sigma_{\mathrm{ul}} \propto \delta\,N^{-1/4}\,.
\end{equation}
In \figurename~\ref{fig:gaussdisp}, I test this analytic estimate with
direct mock-data simulations and inferences. This figure demonstrates
that the large-$N$ asymptotic behavior occurs above about 100 data
points, with a steeper dependence on $N$ between 10 and 100 data
points. The latter is the relevant regime for the data in this paper.

Thus, in the limit of many data points, it is difficult to
significantly improve upon the upper limit on the intrinsic scatter by
observing more stars, especially given the limited number of stars
suitable for high-resolution spectroscopy in all but the nearest
clusters. As \figurename~\ref{fig:gaussdisp} illustrates, if one were
to use the simplistic ``0.1\dex'' standard abundance uncertainty,
about 100,000 stars would be required to limit the intrinsic
dispersion to below $0.01\dex$. It is therefore of the utmost
importance to characterize, understand, and use one's abundance
precision.

The analytic estimate in this section and the simulations in
\figurename~\ref{fig:gaussdisp} also show that given a fixed amount of
observing time $T$, it is more efficient to observe a small number of
stars for longer times (uncertainties and the intrinsic-scatter limit
decrease as $T^{-1/2}$) than to observe a large number of stars for
short times (the intrinsic-scatter limit decreases as $T^{-1/4}$ if $N
\propto T$), at least in the regime where the abundance uncertainties
are limited by photon noise. Thus, higher signal-to-noise ratio
observations of a smaller sample are more important.

\subsection{Overview}

\begin{figure*}
\begin{center}
  \includegraphics[width=0.98\textwidth,clip=]{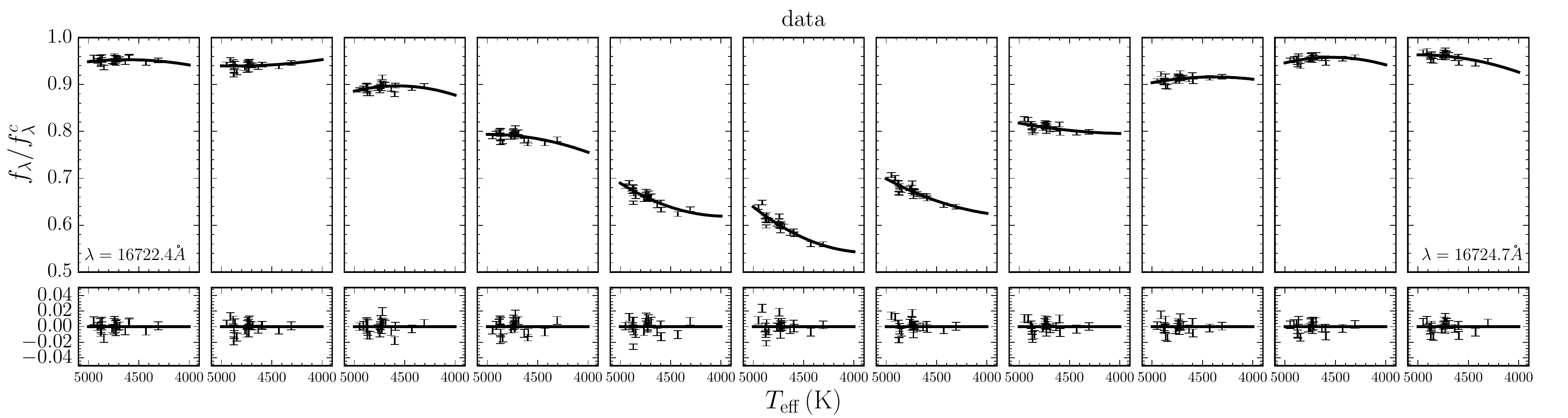}\\
  \includegraphics[width=0.98\textwidth,clip=]{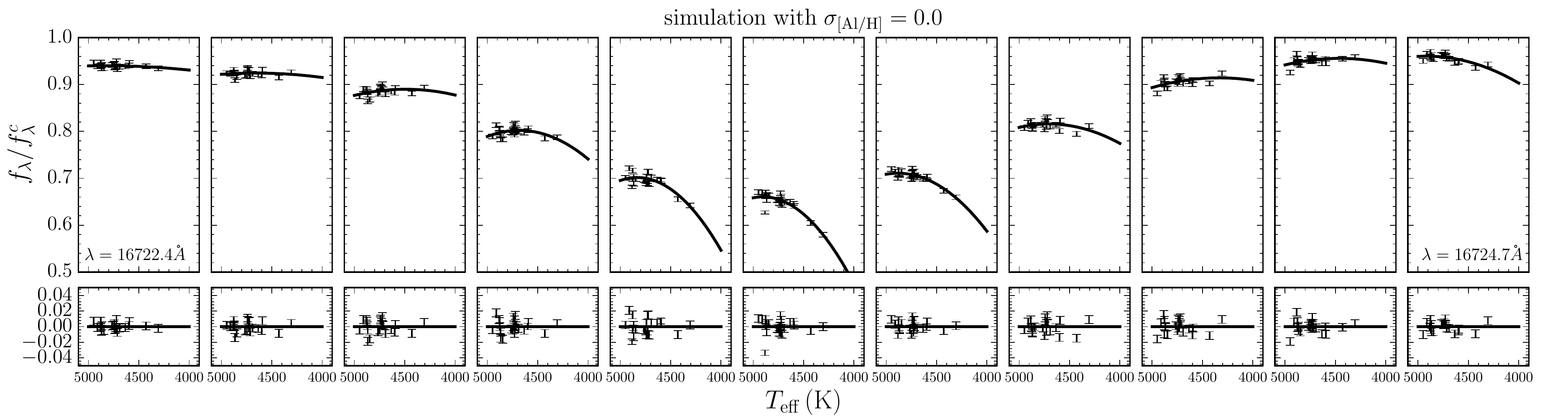}\\
  \includegraphics[width=0.98\textwidth,clip=]{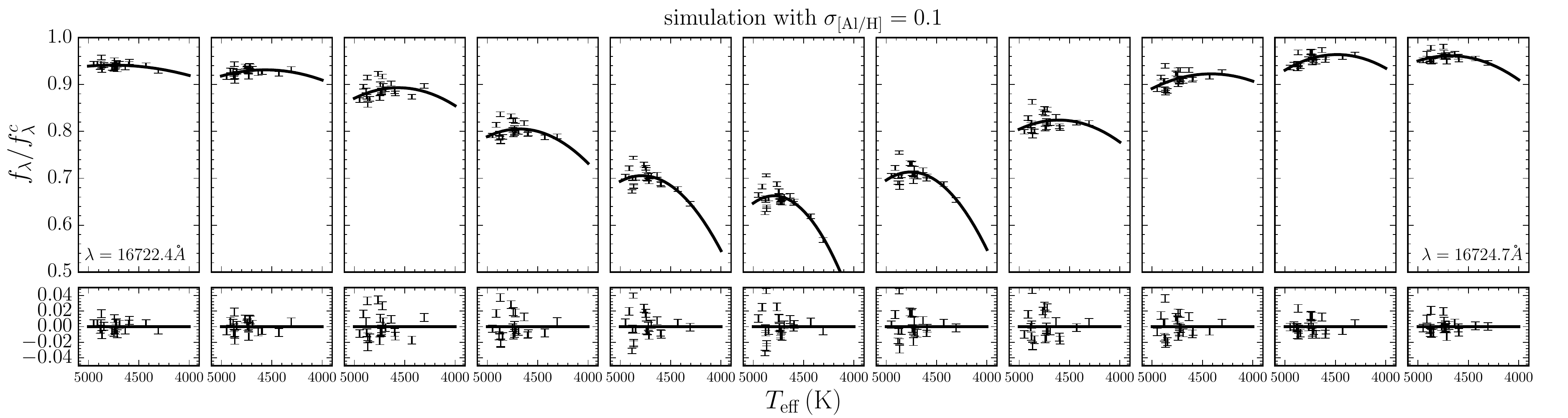}
\end{center}
\caption{The Al spectral line at 16723.5\,\AA\ for giants in M67
  (top). Each column displays the dependence of a single pixel of the
  continuum-normalized spectrum on \teff\ and a quadratic fit to this
  dependence. The bottom panels show the residuals from the fit; these
  are largely consistent with the reported uncertainties. The bottom
  two sub-figures show the same for two simulations of the data: The
  middle figure assumes no intrinsic scatter in the Al abundance and
  the bottom figure uses a scatter of $0.1\dex$. For the latter
  simulation, the residuals are clearly larger than the uncertainties,
  demonstrating that the scatter around the quadratic fit is strongly
  constraining for the intrinsic abundance
  scatter.}\label{fig:simpixal}
\end{figure*}

Motivated by the discussion in the previous section that the abundance
precision is of the highest importance in studying abundance scatter,
and by the fact that data uncertainties are simplest in the space of
the spectra themselves rather than in that of the measured abundances,
I propose a novel method here for limiting the abundance scatter in
clusters and for evaluating similarity and dissimilarity of abundances
in groups of stars more generally. This method evaluates the effect of
abundance scatter through forward modeling on the observed stellar
spectra where it can be directly compared to the spectral
uncertainties. By empirically removing temperature trends in both the
observed and simulated spectra, this method is robust to modeling
errors related to standard assumptions in stellar spectroscopy and to
the main effects of stellar evolution on the surface abundances.

This paper is organized as follows. In \sectionname~\ref{sec:empmodel}
I propose that a simple empirical model for an open cluster without
intrinsic abundance scatter is that all stellar properties are a
one-dimensional function of (fundamentally) initial mass and I discuss
how we can easily test this model. \sectionname~\ref{sec:data}
introduces the APOGEE data for the four open clusters that the method
in this paper is applied to. In \sectionname~\ref{sec:oned}, I
explicitly test the one-dimensional hypothesis for the APOGEE
clusters. I demonstrate that all clusters are consistent with this
hypothesis and that this is strongly constraining for the intrinsic
scatter in 15 elements that have absorption features in the APOGEE
wavelength range. To make the intrinsic-scatter limits precise,
\sectionname~\ref{sec:inference} describes a method using Approximate
Bayesian Computation (ABC) that uses detailed simulations of the
observed spectra to put robust limits on the intrinsic abundance
scatter. I discuss the results, their implications, and future
prospects in \sectionname~\ref{sec:conclusions}.

\appendixname~\ref{sec:clusterspec} displays the high signal-to-noise
``stacked'' red-giant spectra for M67, NGC 6819, and NGC 2420 that I
obtain using the method in \sectionname~\ref{sec:empmodel}. The
remaining three appendices discuss relevant technical details of the
observed and synthetic APOGEE spectra. \appendixname~\ref{sec:repeats}
discusses how I construct an empirical error model for the APOGEE
continuum-normalized spectra using repeat observations of
stars. \appendixname~\ref{sec:synspec} provides details of the
procedure and code to generate synthetic APOGEE spectra tailored to
each star for variations of all 15 considered
elements. \appendixname~\ref{sec:windows} presents an investigation of
the sensitivity of APOGEE spectra to abundance changes of different
elements; these sensitivities are a crucial ingredient in constraining
the abundance scatter using forward simulations.

\section{An empirical model for the spectra of open-cluster members}\label{sec:empmodel}

The model for open clusters that we are interested in constraining is
that they consist of a set of stars born from a well-mixed gas cloud
in a negligible amount of time ($\lesssim10\Myr$). Assuming no scatter
in the birth abundances, the most important factor distinguishing
different stars is their different initial mass, which spans the range
$\approx0.1\msun$ to $\approx100\msun$. Each star's initial mass
together with the common initial abundances determines its subsequent
evolution. At the present day, we observe stars in clusters to span a
wide range of luminosities, temperatures, surface gravities, etc. due
to the range in initial masses. Every photometric and spectroscopic
property of cluster stars should then follow a one-dimensional
relation as a function of stellar mass. In particular, the spectra of
cluster stars near absorption features of different elements should
follow a one-dimensional sequence. Without any reference to particular
models of stellar photospheres, this is a testable prediction.

An important advantage of this approach to testing the chemical
homogeneity of open clusters, is that many of the stellar evolution
effects on the surface abundances that normally confound studies of
homogeneity by (correctly) showing abundance scatter in the
\emph{current} abundances \citep[\eg,][]{Onehag14a}, are themselves
primarily functions of the stellar mass (e.g., gravitational settling
or mixing of C, N, and O during dredge-up episodes). Similarly,
hydrodynamical effects often parametrized in simplified treatments
using micro- and macroturbulence are mostly functions of the current
evolutionary state (temperature, gravity) and therefore also functions
of stellar mass. Theoretically predicting these functions is
difficult, but it is clear that the combination remains
one-dimensional and fitting a flexible one-dimensional model allows us
to ignore this lack of knowledge.

Besides these deterministic effects, random effects in the initial
condition of each star or in its subsequent evolution can break the
one-dimensional model. For example, stars are born with a distribution
of initial rotation speeds. When these survive to the present time,
they will give rise to different line profiles. The effects of
magnetic braking likely cause all stars in a cluster to have the same
current rotation for the old clusters that we are interested in here
($>1\Gyr$; \citealt{Weber67a,Skumanich72a,Kawaler88a}), but a random
distribution of inclinations will still give rise to different line
profiles. For the giants that I consider in this paper, the line
broadening due to rotation can, however, be neglected. Differences in
initial rotation might additionally induce variations in the internal
mixing that could manifest themselves at the surface
today. Interactions between binary stars may also lead to abundance
anomalies that would be uncorrelated with mass, or the infall of rocky
planets could lead to scatter in the abundances of refractory elements
\citep{Melendezz09a}. While this might confound studies of the initial
chemical homogeneity of clusters, limits on the abundance scatter
obviously constrain the importance of these processes.

Beyond effects intrinsic to the cluster stars, instrumental effects
may give rise to scatter at fixed initial stellar mass, even when all
stars are observed with the same instrument. Foremost among these are
variations in the line-spread function (LSF), which lead to different
broadening profiles similar to the case of rotation discussed
above. For the APOGEE spectra that I employ in this paper, LSF
variations exist, but are small enough that they are only confused
with abundance scatter $\lesssim0.01\dex$. In the forward modeling
approach below, I take the LSF variations among cluster stars fully
into account.

The basic model that I will use in this paper is therefore that the
spectra of stars in open clusters are a one-dimensional
sequence. Because (initial) stellar mass is difficult to observe, I
will employ the effective temperature \teff\ as a proxy for the mass
and use one-dimensional models as a function of temperature. For the
red giants that we will consider later, \teff\ is a good proxy for the
mass\footnote{As we will see, this does not hold exactly because of
  the presence of both red-giant and red-clump stars at temperatures
  $\approx4750\,\mathrm{K}$. Red-clump stars have slightly different
  surface gravities at the same \teff\ and may also have different C
  and N abundances due to the effect of convective mixing on the upper
  giant branch.} and we have photometric \teff\ available that are
independent of the considered spectra. Each pixel value $f_\lambda^i$
for different stars $i=1\ldots N$ in a cluster and wavelengths
$\lambda$ can then be modeled as a function
$g_\lambda(\teff|\theta_\lambda)$ characterized by parameters
$\theta_\lambda$ plus the measurement noise
\begin{equation}
  f_\lambda^i = g_\lambda(\teffi|\theta_\lambda) + \mathrm{noise}\,.
\end{equation}
I model $g_\lambda$ as a second-order polynomial in \teff\ and we can
then fit for the parameters $\theta_\lambda$ at each wavelength using
the observed $f_\lambda^i$ and their catalog uncertainties using
maximum likelihood; I also include an intrinsic scatter in the fit,
but this is always small. This approach is similar to that taken by
\citet{Ness15a} for deriving an empirical model of stellar spectra
using a calibration sample that can then be applied to determine
stellar parameters (\teff\ in this case). The approach taken here is
different in that I fit the empirical model only as a way of
determining whether the spectra of stars in a cluster are the same
apart from trends with \teff.

Once we have determined the best-fitting
$g_\lambda(\teffi|\theta_\lambda)$ for each pixel $\lambda$, we can
compute the residuals which are given by
\begin{equation}
  r_\lambda^i = f_\lambda^i-g_\lambda(\teffi|\theta_\lambda)\,.
\end{equation}
If the one-dimensional model provides a good fit, then the
distribution of residuals should be consistent with the uncertainty
distribution for each pixel.

This procedure is illustrated in \figurename~\ref{fig:simpixal}. The
`data' row of this figure displays APOGEE spectra for stars in M67
(described in more detail below) in the region of an Al feature. Each
column sub-panel shows the dependence of the observed flux on
\teff\ and the polynomial fit to this dependence. The bottom panels
display the residuals from the fit and it is clear by eye that the
residuals are largely consistent with the reported uncertainties.

\section{Data}\label{sec:data}

\begin{figure}
  \includegraphics[width=0.48\textwidth,clip=]{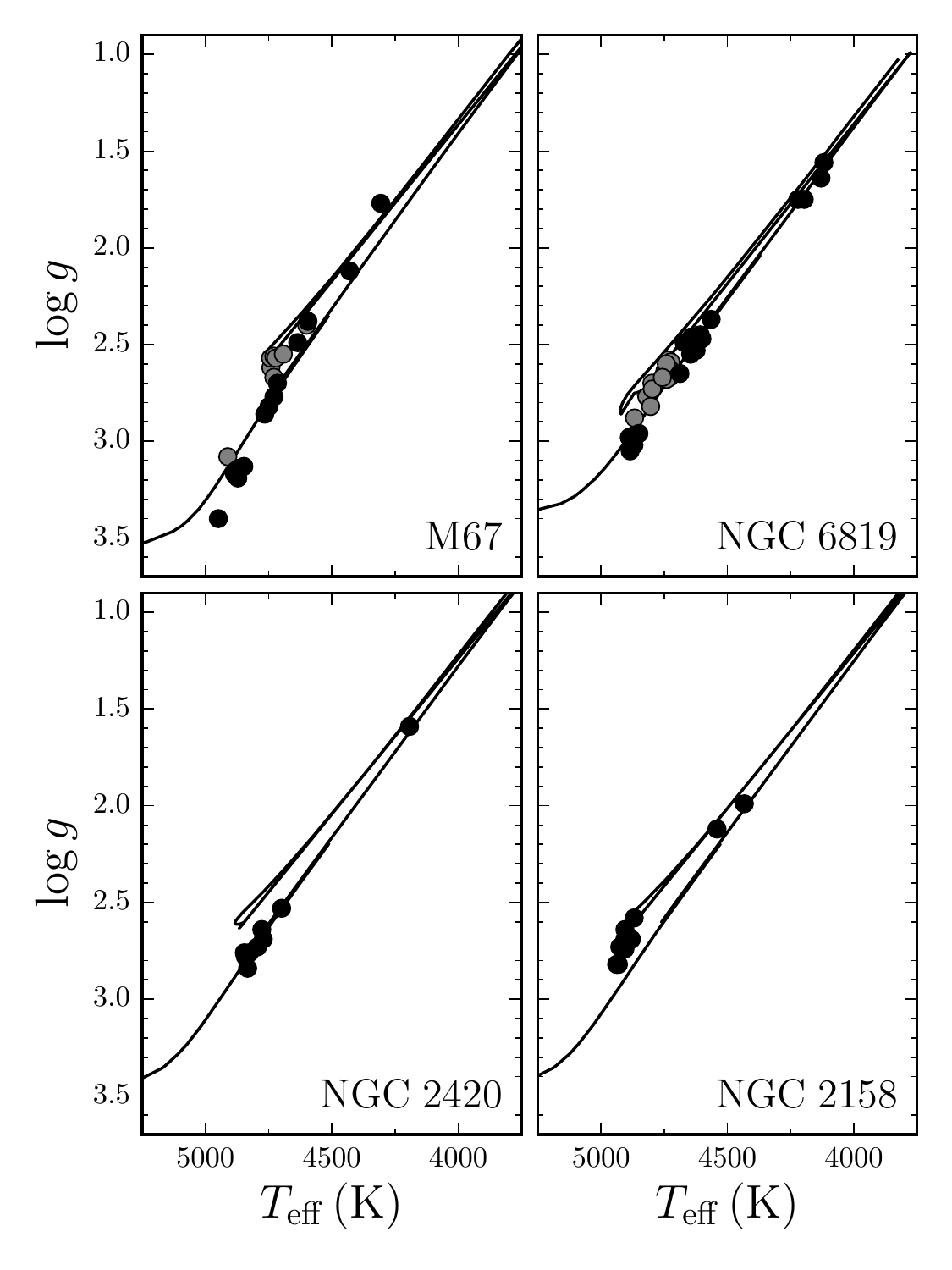}
\caption{\logg\ vs. \teff\ for the four open clusters studied in this
  paper. Likely red-clump stars have been colored gray for M67 and NGC
  6819. These are excluded for the analysis of C and N for M67 and are
  altogether excluded for the analysis of NGC 6819. PARSEC isochrones
  \citep{Bressan12a} for ages of 2.5, 1.6, 2, and 2\Gyr\ at the
  cluster's metallicity are shown as well.}\label{fig:spechr}
\end{figure}

The spectroscopic data for open-cluster members that I analyze here
comes from the SDSS-III/\apogee\ \citep{Majewski15a}, a
high-resolution ($R\approx 22,500$) spectroscopic survey that observes
in the $H$-band (1.51 to 1.70 $\mu$m) using a 300-fiber spectrograph
(\citealt{Wilson10a}, J.~Wilson \etal\ 2015, in preparation) on the
2.5-meter Sloan Foundation telescope \citep{Gunn06a}. I use data from
the public Data Release 12 \citep[DR12;][]{Alam15a,Holtzman15a} for
four open clusters that have a large number of members: M67
($\feh\approx0$), NGC 6819 ($\feh\approx0$), NGC 2158
($\feh\approx-0.15$), and NGC 2420 ($\feh\approx-0.2$). Members for
these clusters are obtained from the catalog in \citet{Meszaros13a}
and I use the photometric \teff, surface gravities \logg, and overall
metallicities \feh\ used in that paper. Full details on the APOGEE
target selection for these clusters can be found in
\citet{Zasowski13a}. I only select the stars with $4000\,\mathrm{K}
\leq \teff \leq 5000\,\mathrm{K}$, because this range contains most of
the stars and spectral modeling is more uncertain at $\teff <
4000\,\mathrm{K}$. After this cut, the sample consists of 24 stars, 8
of which are likely red-clump stars, for M67; 30 stars, 13 of which
are likely red-clump stars, for NGC 6819; 10 stars in NGC 2158; and 9
stars in NGC 2420. The \logg\ vs. \teff\ for the giants in these
clusters are displayed in \figurename~\ref{fig:spechr}.

The spectra used here are the \texttt{apStar} spectra that combine
data from all APOGEE observations of a given star; I use the version
using a ``global'' weighting of the individual spectra going into the
combination. These spectra are sampled on a logarithmic wavelength
grid in the restframe of the star \citep{Nidever15a}. In
\appendixname~\ref{sec:repeats}, I also use the spectra from the
individual hour-long APOGEE observations that are combined to form the
\texttt{apStar} spectra; these are also available from the SDSS data
base on the same wavelength grid. Pixels identified as bad in the
\texttt{APOGEE\_PIXMASK} bitmask for the reasons \texttt{BADPIX},
\texttt{CRPIX}, \texttt{SATPIX}, \texttt{UNFIXABLE}, \texttt{BADDARK},
\texttt{BADFLAT}, \texttt{BADERR}, \texttt{NOSKY}, or
\texttt{SIG\_SKYLINE} \citep{Holtzman15a} are given very large
uncertainties to remove them from further consideration. Uncertainties
smaller than $0.5\,\%$ are set to $0.5\,\%$, because of systematic
errors at the $0.5\,\%$ level for APOGEE spectra \citep{Nidever15a}.

I fit for the continuum of each spectrum using the method of
\citet{Ness15a}. This method identifies a set of continuum pixels by
fitting a quadratic model in $(\teff,\logg,\feh,\afe)$ to each pixel
for a calibration sample with known $(\teff,\logg,\feh,\afe)$ and
selecting those pixels whose values display only a small dependence on
$(\teff,\logg,\feh,\afe)$. I use the same calibration sample as used
in \citet{Ness15a} and select pixels with linear dependencies less
than $(10^{-5}\,\mathrm{K},0.006,0.012,0.03)$ in
$(\teff,\logg,\feh,\afe)$ (similar to \citealt{Ness15a}) and
additionally limit the pixels to those with intrinsic scatter less
than $0.015$ to remove pixels with large variations that cannot be
attributed to the basic stellar parameters. Using the wavelengths of
these continuum pixels, the continuum for each star is determined by
fitting a fourth-order polynomial over the wavelength range of each of
the three APOGEE detectors to just these wavelengths. As demonstrated
in \appendixname~\ref{sec:repeats} using repeat observations, this
procedure is highly stable and produces consistent
continuum-normalized spectra for different observations of the same
star. After continuum normalization, uncertainties smaller than
$0.005$ are set to $0.005$ for the same reason as above. I further
remove pixels with errors larger than 0.02 (signal-to-noise ratio $<
50$) from further consideration, because the errors for these low
signal-to-noise ratio pixels might not be well-characterized by the
reported uncertainty (this includes all of the pixels flagged as bad
mentioned earlier). This only removes a few percent of the pixels.

Ideally, the errors in the spectra should be well-characterized by the
reported uncertainties in the APOGEE database, which assume
uncorrelated errors between different pixels. I test the reported
uncertainties in \appendixname~\ref{sec:repeats} using 4,143 repeat
observations of 1,381 stars bright enough that each individual
hour-long exposure has high signal-to-noise ratio. These tests
demonstrate that the reported uncertainties are typically
underestimated by 10 to 20\,\%, but ranging up to 100\,\% for
significant portions of the wavelength range, especially near the
ubiquitous telluric absorption features. Furthermore, these tests show
that errors display significant correlations out to dozens of pixels
($\gtrsim10\,\AA$). This is a range that is almost ten times as wide
as the reported LSF. This large range over which correlations are
significant is most likely due to correlated errors induced by the
continuum normalization. In what follows I use the residuals from
repeat observations directly as an empirical sampling of the
uncertainty in the observed spectra (see further discussion in
\appendixname~\ref{sec:repeats}).

\begin{deluxetable}{lcccc}
  \tablecaption{Expected abundance uncertainties}
  \tablecolumns{2}
  \tablewidth{0pt}
  \tablehead{\colhead{Element} & \colhead{M67} & \colhead{NGC 6819} & \colhead{NGC 2420}}
  \startdata
  C  & 0.04 (0.03--0.07) & 0.05 (0.02--0.07) & 0.06 (0.02--0.07)\\
N  & 0.07 (0.06--0.09) & 0.06 (0.04--0.08) & 0.08 (0.05--0.08)\\
O  & 0.16 (0.04--0.20) & 0.13 (0.04--0.20) & 0.16 (0.03--0.20)\\
Na  & 0.13 (0.08--0.17) & 0.09 (0.06--0.12) & 0.16 (0.09--0.20)\\
Mg  & 0.04 (0.03--0.05) & 0.03 (0.03--0.04) & 0.03 (0.03--0.04)\\
Al  & 0.09 (0.08--0.10) & 0.04 (0.04--0.04) & 0.04 (0.04--0.05)\\
Si  & 0.06 (0.06--0.06) & 0.04 (0.04--0.05) & 0.04 (0.04--0.05)\\
S  & 0.12 (0.11--0.20) & 0.07 (0.07--0.09) & 0.08 (0.08--0.11)\\
K  & 0.06 (0.06--0.07) & 0.04 (0.03--0.04) & 0.04 (0.04--0.04)\\
Ca  & 0.05 (0.04--0.05) & 0.04 (0.04--0.04) & 0.04 (0.04--0.04)\\
Ti  & 0.07 (0.05--0.10) & 0.06 (0.04--0.09) & 0.10 (0.04--0.12)\\
V  & 0.08 (0.03--0.14) & 0.08 (0.03--0.13) & 0.15 (0.04--0.20)\\
Mn  & 0.06 (0.05--0.07) & 0.03 (0.03--0.04) & 0.06 (0.04--0.06)\\
Fe  & 0.06 (0.05--0.06) & 0.04 (0.04--0.04) & 0.04 (0.04--0.04)\\
Ni  & 0.07 (0.07--0.07) & 0.06 (0.05--0.06) & 0.06 (0.06--0.06)\\

  \enddata
  \tablecomments{Expected abundance uncertainties in the abundance of
    X computed from $\Delta \chi^2(\mathrm{X})$ (weighted using the
    pixel weights of \appendixname~\ref{sec:windows}) of a baseline
    model with solar abundance ratios for each cluster star. The
    median precision for each element and each cluster is shown, as
    well as the full range of all the cluster members in
    parenthesis. These precisions assume perfect spectral models,
    perfect knowledge of all other stellar parameters and abundances,
    and that the APOGEE noise model is correct, but by using the pixel
    weights they only use parts of the spectrum that are sensitive to
    each element. Therefore, these represent a realistic estimate of
    the expected precision that can be hoped to be
    achieved.}\label{table:abuprec}
\end{deluxetable}

\appendixname~\ref{sec:synspec} discusses the details of how I
generate synthetic APOGEE spectra for each cluster star individually
using its $(\teff,\logg)$, the median cluster metallicity, and
variations in the abundances of individual elements. Using these
synthetic spectra, we can estimate the precision with which we can
measure the abundances of individual elements by computing the $\Delta
\chi^2([\mathrm{X/H}])$ from a baseline model where all abundance
ratios are solar. I compute this $\chi^2([\mathrm{X/H}])$ weighting
the contribution of each pixel with the pixel-weights that give
prominence to clean absorption features of each element (see
\appendixname~\ref{sec:windows}). Assuming perfect knowledge of all
other parameters (\teff, \logg, \feh, but also micro- and
macroturbulence, etc.) and that our modeling is perfect, this
effectively sets a realistic lower limit on the precision. I compute
$\chi^2([\mathrm{X/H}])$ using the reported uncertainties (\ie, not
taking into the account the underestimation of the uncertainties),
which also makes the estimated abundance uncertainties a lower
limit. These estimated abundance uncertainties are shown in
\tablename~\ref{table:abuprec}. From \figurename~\ref{fig:gaussdisp},
we expect the $95\,\%$ upper limit on the abundance scatter in each
individual cluster to be roughly the precision of an individual
abundance measurement (slightly smaller for M67 and NGC 6819, slightly
larger for NGC 2420). The precision for some elements (like C, N, and
O) sensitively depends on temperatures, leading to a wide range of
expected precision. If we combine all clusters, the $\sim\!50$ stars
should give a 95\,\% upper limit that is about $60\,\%$ of the
abundance precision. It is therefore clear that we should be able to
extract limits on the intrinsic abundance scatter for most of these
elements at the level of a few times $0.01\dex$.

\section{Are open clusters one-dimensional sequences?}\label{sec:oned}

\begin{figure*}[t]
  \begin{center}
  \includegraphics[width=0.88\textwidth,clip=]{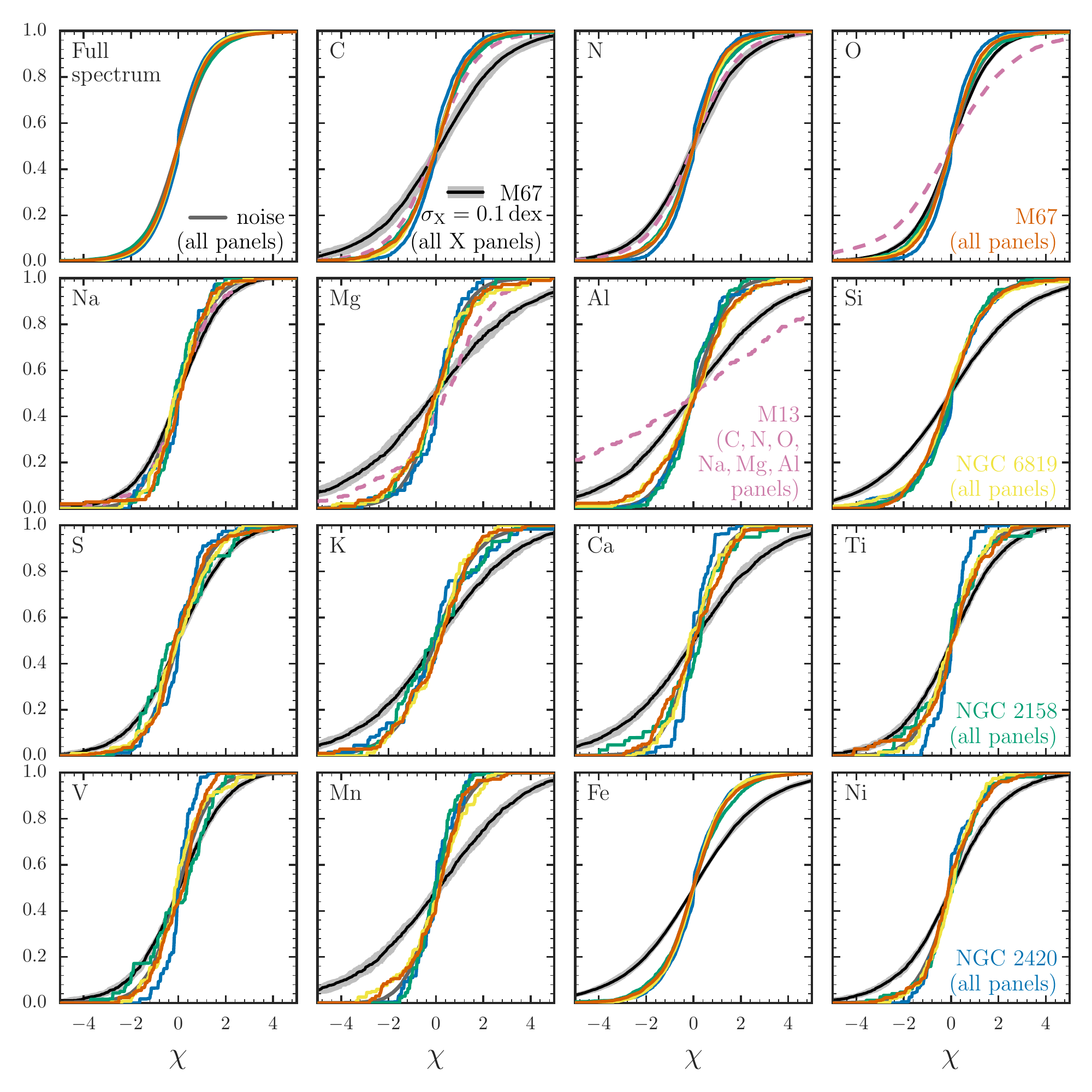}
  \end{center}
\caption{Cumulative distribution of the normalized residuals from the
  quadratic \teff\ fit to each pixel of the continuum-normalized
  spectra for members of the open clusters M67, NGC 6819, NGC 2158,
  and NGC 2420. The top, left panel shows the cumulative distribution
  of all pixels in the spectrum. The other panels display the same
  residuals, but weighted using the pixel weights that give prominence
  to the pixels most affected by a given element (see
  \appendixname~\ref{sec:windows}). The color-coding of the different
  clusters is indicated by the text labels. The dark gray line shows
  the distribution of normalized residuals due to errors in the
  spectra derived from repeat observations (see
  \appendixname~\ref{sec:repeats}); this line cannot be clearly seen
  in most panels because it lies beneath the curves for the four open
  clusters. The dashed line shows the residuals computed using a
  similar quadratic fit for M13, a globular cluster with known scatter
  and anti-correlations in the abundances of light elements (these are
  only shown for C, N, O, Na, Mg, and Al). The black line and light
  gray band give the median and interquartile range of simulated data
  for M67 with an abundance spread of $\sigma_{\xh} = 0.1\dex$. The
  distribution of residuals around the one-dimensional \teff\ fit for
  all clusters is consistent with that expected from the random errors
  in the spectra for all elements. An abundance scatter of $0.1\dex$
  in any of the 15 elements considered here would give rise to much
  larger residuals.}\label{fig:cumul}
\end{figure*}

To check whether or not the four open clusters described in
\sectionname~\ref{sec:data} are consistent with forming a
one-dimensional temperature (as a proxy for mass) sequence, I fit the
quadratic model to the \teff\ dependence of each pixel using members
in each cluster as discussed in \sectionname~\ref{sec:empmodel}. M67
and NGC 6819 both have a significant number of both red-clump and
first-ascent red-giant stars at similar temperatures (see
\figurename~\ref{fig:spechr}). These lead to two potential issues with
the method used in this paper. Firstly, deep mixing along the upper
giant branch \citep[\eg,][]{Gilroy91a} can change the surface C and N
abundances. While this should be a smooth change as a function of
temperature, when stars move to the red-clump after the helium flash,
as a function of temperature alone this leads to stars in the cluster
having a bimodal C and N distribution that cannot be captured with the
model from \sectionname~\ref{sec:empmodel}. Secondly, red-clump stars
have slightly lower \logg\ than first-ascent red-giant stars at the
same \teff. This again leads to variations in the spectra at a fixed
\teff\ that are not included in our model. To avoid these issues, I
remove all red-clump stars in M67 when looking at C and N, but not
when considering the other elements. The number of red-clump giants is
large enough in NGC 6819 that their \logg\ differences lead to
substantial scatter and I therefore remove all clump giants altogether
in NGC 6819. If we were to fit the spectra as a function of mass
rather than \teff, these issues would be avoided, but we currently do
not have precise enough masses to do this. In NGC 6819, I also remove
the Li-rich giant 2M19411367+4003382, which may not be a cluster
member or has anomalously low mass if it is \citep{Carlberg15a}. The
final sample for NGC 6819 therefore consists of 16 stars.

I further find through visual inspection that the spectra of stars in
NGC 2158 have significant issues with continuum normalization due to a
large fraction of bad pixels. The regions with bad continuum
normalization are identified by eye and removed from further
consideration. The remaining data is for so few stars and so few
pixels that no interesting constraints on the abundance scatter can be
placed. I discuss the results for NGC 2158 in this section, but do not
consider it further.

Thus, I compute the residuals from the quadratic fit for all pixels
(see \figurename~\ref{fig:simpixal} for an illustration of the
quadratic fit and its residuals for pixels near an Al line) and
normalize them using the reported pixel-level uncertainties. The
cumulative distribution of all normalized residuals for all four
clusters are displayed in the top, left panel of
\figurename~\ref{fig:cumul}. They are compared with the distribution
of normalized residuals from repeat observations, which give the
distribution expected from random errors in the spectra alone (see
\appendixname~\ref{sec:repeats}). These noise residuals have a very
similar distribution to those from the quadratic fit; the noise line
cannot be seen because it lies underneath those from the four
clusters.

The other panels in \figurename~\ref{fig:cumul} show the same
residuals, but weighted using the pixel-level weights for different
elements described in \appendixname~\ref{sec:windows}. These weights
for a given element essentially correspond to the derivatives of model
spectra with respect to the abundance of that element. Thus, these
panels display the residuals in regions dominated by the effects of a
given element, giving higher weight to those pixels which are most
strongly affected by changes in that element. We see that the
distribution of fit residuals is consistent with that of the noise
residuals for practically all elements and all clusters. For
comparison, I have also performed the same kind of fit for 62 members
of M13, a globular cluster that displays significant anti-correlations
in its light-element abundances (see \citealt{Meszaros15a} for a study
of this using the APOGEE data). The distribution of residuals for M13
is shown for the light elements (atomic number $<14$). It is clear
that a significant dispersion in C, N, O, Mg, and Al is present for
M13 (that for Na cannot be detected, because the Na line used here is
too weak at the metallicity of M13).

Thus, we see that the spectra of members of all four open clusters are
consistent to within their uncertainties with forming a
one-dimensional function of \teff. I do not attempt to quantify this
consistency further here, but instead directly infer constraints on
the abundance scatter using a similar method in
\sectionname~\ref{sec:inference}. To get a sense of how strong a
constraint on the abundance scatter of different elements this
consistency implies, I have computed 100 sets of mock spectra for all
M67 stars assuming an intrinsic scatter of $\sigma_{\xh} = 0.1\dex$
using the procedure described in \appendixname~\ref{sec:synspec} and
have fit each set with the same quadratic \teff\ model as the real
data. An example simulation is displayed in the bottom panel of
\figurename~\ref{fig:simpixal}. The median and interquartile range of
these 100 simulations for each element is shown in
\figurename~\ref{fig:cumul}. For all elements a scatter of $0.1\dex$
would give a distribution of the normalized residuals that is much
wider than the observed distribution for M67 and that is much wider
than can be explained by the errors in the spectra.

\begin{figure*}[t]
  \begin{center}
  \includegraphics[height=0.23\textheight,clip=]{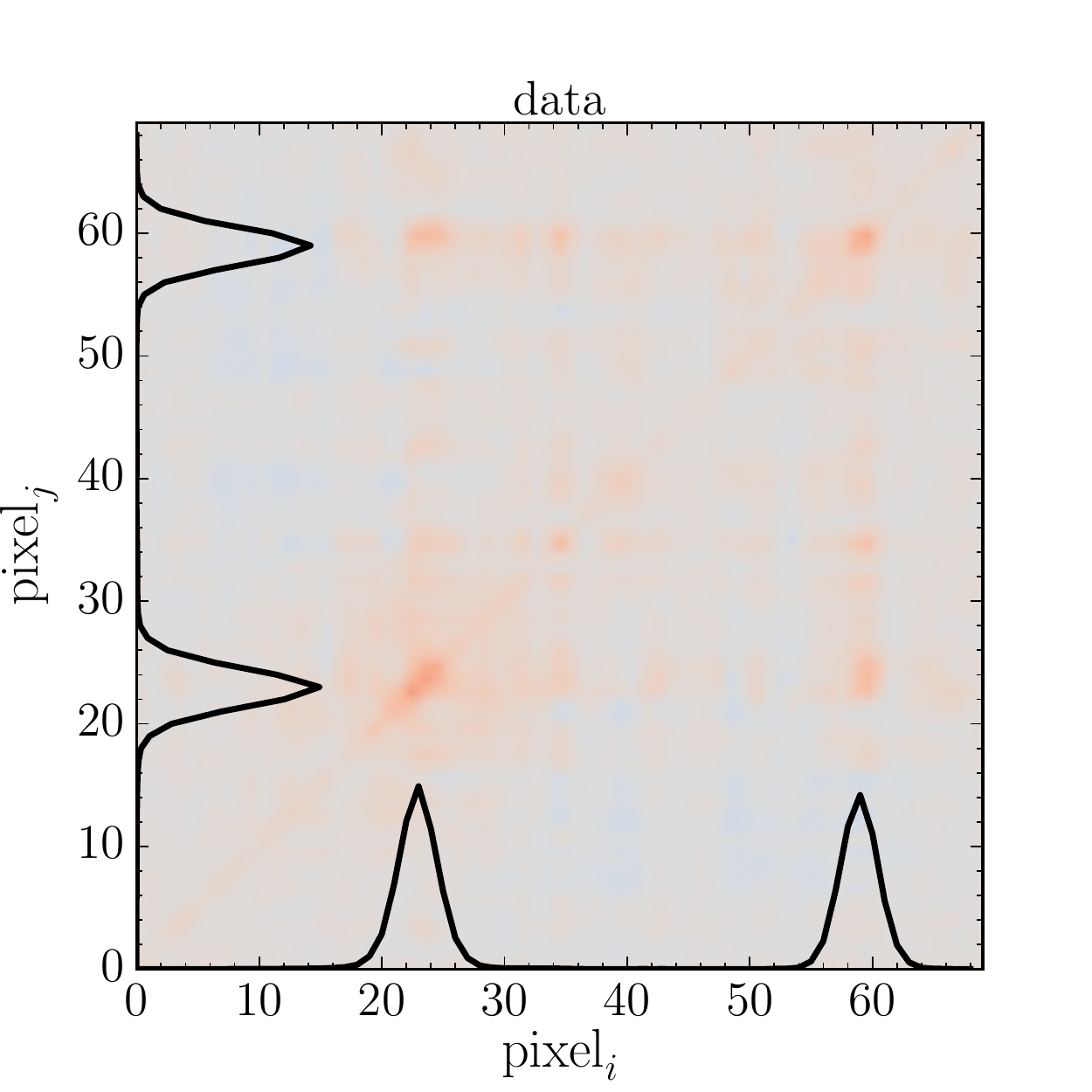}\kern-1.5em
  \includegraphics[height=0.23\textheight,clip=]{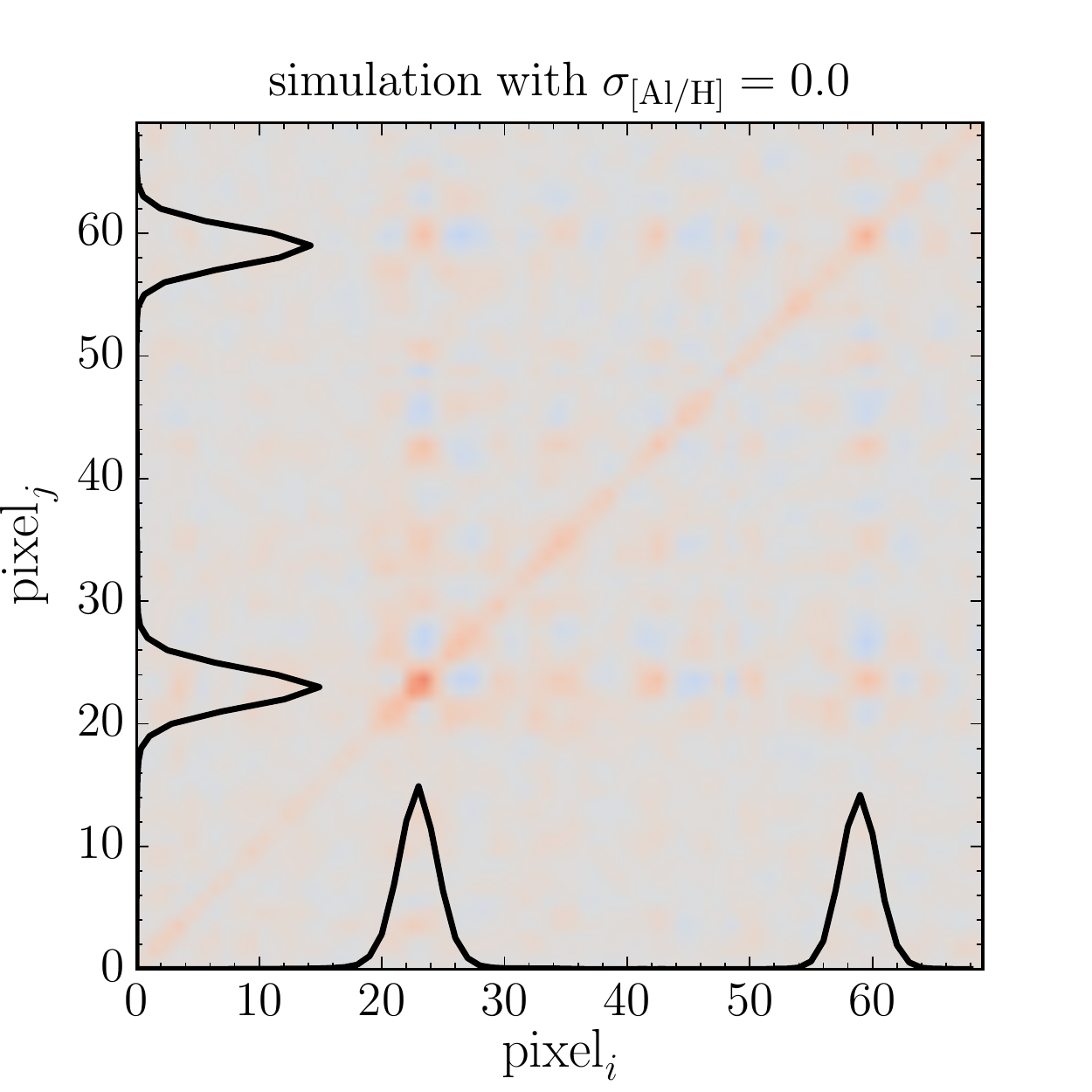}\kern-1.8em
  \includegraphics[height=0.23\textheight,clip=]{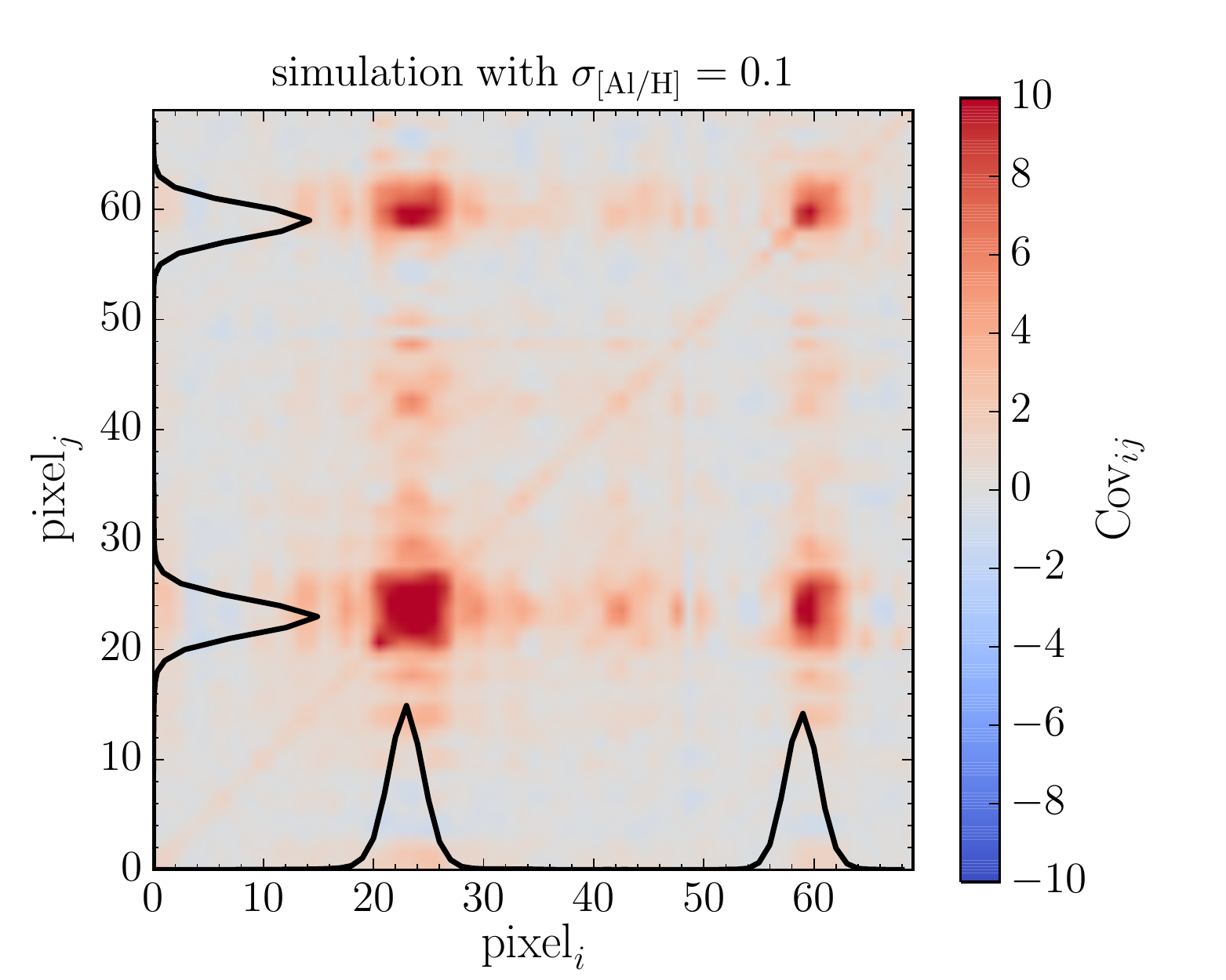}
\caption{Covariance matrix of the normalized residuals of the data for
  members of M67 (left panel) and for the simulations assuming
  $\sigma_{[\mathrm{Al/H}]} = 0$ (middle panel) and
  $\sigma_{[\mathrm{Al/H}]} = 0.1\dex$ from
  \figurename~\ref{fig:simpixal}. Only the 69 pixels with non-zero Al
  weights are included. The Al weights are displayed as the black
  lines on both the $x$ and $y$ axis; these peak at the two Al lines
  used here. Intrinsic abundance scatter in Al leads to a large
  scatter in the residuals near the Al line and to large correlations
  between the residuals at different absorption wavelengths that can
  be used to constrain the intrinsic abundance
  scatter.}\label{fig:simpixal_cov}
\end{center}
\end{figure*}

As the spectra of all cluster members are consistent with being the
same except for their temperatures, we can use the quadratic fit to
construct a high signal-to-noise ratio spectrum for the cluster at a
given temperature. This combined \emph{cluster spectrum} at $\teff =
4750\,\mathrm{K}$ for the red giants in M67, NGC 6819, and NGC 2420
is displayed and discussed in \appendixname~\ref{sec:clusterspec}.

\section{Inference of the abundance scatter}\label{sec:inference}

To determine constraints on the abundance scatter of different
elements for each cluster, I use Approximate Bayesian Computation
(ABC) to construct an approximation of the posterior probability
distribution function (PDF) of the scatter $\sigma_{\xh}$ in each
element X. ABC is an inference technique that approximates the PDF
without explicitly evaluating the likelihood, but instead making use
of forward simulations of the data. To explicitly evaluate the
likelihood of $\sigma_{\xh}$ would require an actual model for the
noise in the spectra---which is difficult to establish (see
\appendixname~\ref{sec:repeats})---and it would be computationally
expensive, because we would need to marginalize over the individual
abundances of each cluster member, while properly taking into account
the varying LSF. However, it is straightforward to generate simulated
data for any $\sigma_{\xh}$ that take into account LSF variations, the
noise and its correlations in the spectra, and that are robust against
systematics in the abundances due to, \eg, deep mixing or deviations
from local thermodynamic equilibrium.

For any $\sigma_{\xh}$, I draw a set of $N$ abundances \xfe\ for the
$N$ cluster members, generate synthetic spectra using the procedure
described in \appendixname~\ref{sec:synspec}, and then fit the
\teff\ dependence of each pixel using the quadratic model described in
\sectionname~\ref{sec:empmodel} in the same way as for the data. An
example of this is displayed for one of the Al lines in
\figurename~\ref{fig:simpixal}. The top panel of that figure shows the
data, while the middle and bottom panels show simulated data with
$\sigma_{\xh} = 0.0\dex$ and $\sigma_{\xh} = 0.1\dex$,
respectively. When running ABC, we retain those $\sigma_{\xh}$ that
lead to similar residuals from the quadratic fit as found in the
data. By only considering a match between the data and the simulated
data in terms of their residuals, we focus the comparison on the
abundance scatter, rather than on whether the simulations produce the
exact same continuum levels, the same line strengths (which may be
affected by such effects as deviations from local thermodynamic
equilibrium or hyperfine structure), the correct behavior of weak and
strong lines of a given element, and whether evolutionary changes in
the surface abundances are included.

\begin{figure}[t!]
\begin{center}
  \includegraphics[width=0.48\textwidth,clip=]{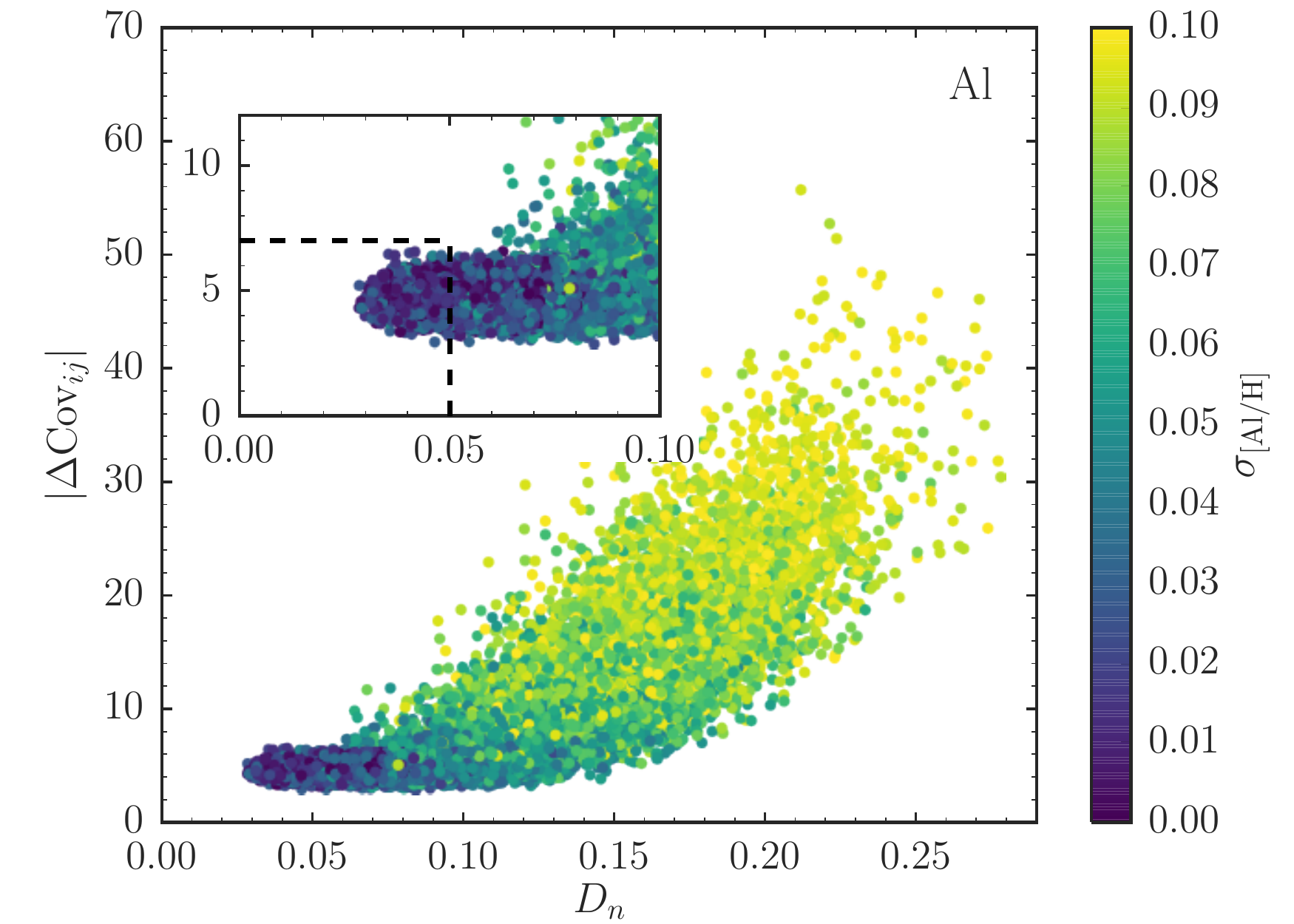}\\
  \includegraphics[width=0.48\textwidth,clip=]{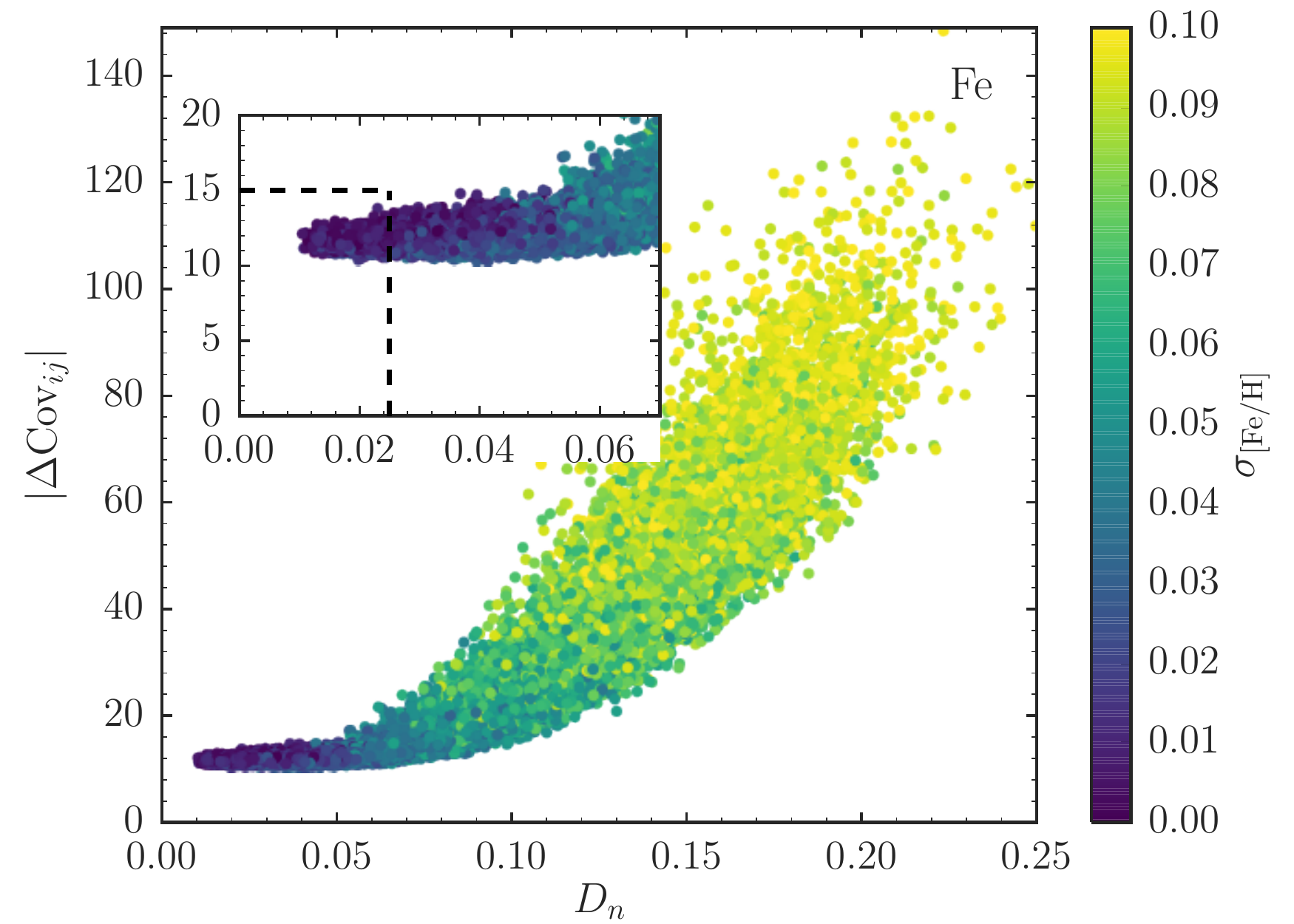}
\end{center}
\caption{Two summary statistics employed in the ABC simulations in
  \sectionname~\ref{sec:inference} to evaluate the similarity between
  the data and the simulated data. The top panel shows simulations
  with intrinsic scatter in Al and the bottom panel has simulations
  with scatter in Fe, both for stars in M67. The $x$ axis has the
  Kolmogorov-Smirnov distance $D_n$ between the cumulative
  distribution of the normalized residuals of the data and the
  simulated data (see \figurename~\ref{fig:cumul}). The $y$ axis has
  the difference $|\Delta \mathrm{Cov}_{ij}|$ between the covariance
  matrix of the residuals of the data and the simulated data. Both
  summary statistics use residuals that are weighted using the weights
  for Al in the top panel and Fe in the bottom panel. The simulations
  are color-coded by their value of $\sigma_{\xh}$. Especially
  $|\Delta \mathrm{Cov}_{ij}|$ is an excellent summary statistic and
  is strongly correlated with $\sigma_{\xh}$, leading to strong
  constraints on $\sigma_{\xh}$. Using the statistic $D_n$ adds
  information on the shape of the distribution of residuals as
  well. The inset zooms in on those simulations that are closest to
  the data and the dashed lines display the cuts used to define the
  final $\sigma_{\xh}$ ABC sampling.}\label{fig:m67stat}
\end{figure}

\begin{figure}[t!]
  \includegraphics[width=0.52\textwidth,clip=]{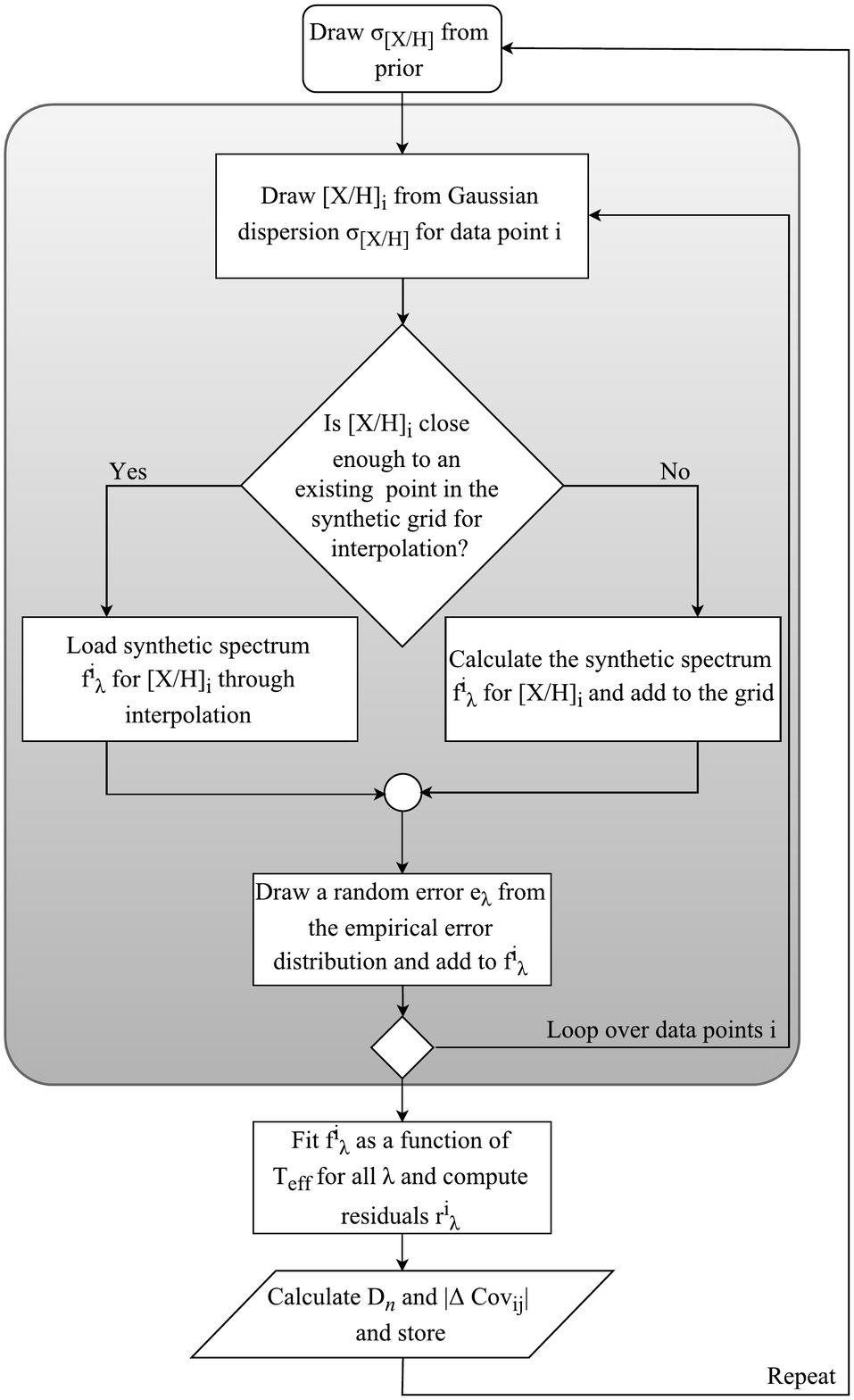}
\caption{Overview of the steps involved in running a single ABC
  simulation. This procedure is repeated until a large number of
  simulations that pass the cuts in $D_n$ and $|\Delta
  \mathrm{Cov}_{ij}|$ are obtained.}\label{fig:flowchart}
\end{figure}

\begin{figure*}
  \begin{center}
  \includegraphics[width=0.88\textwidth,clip=]{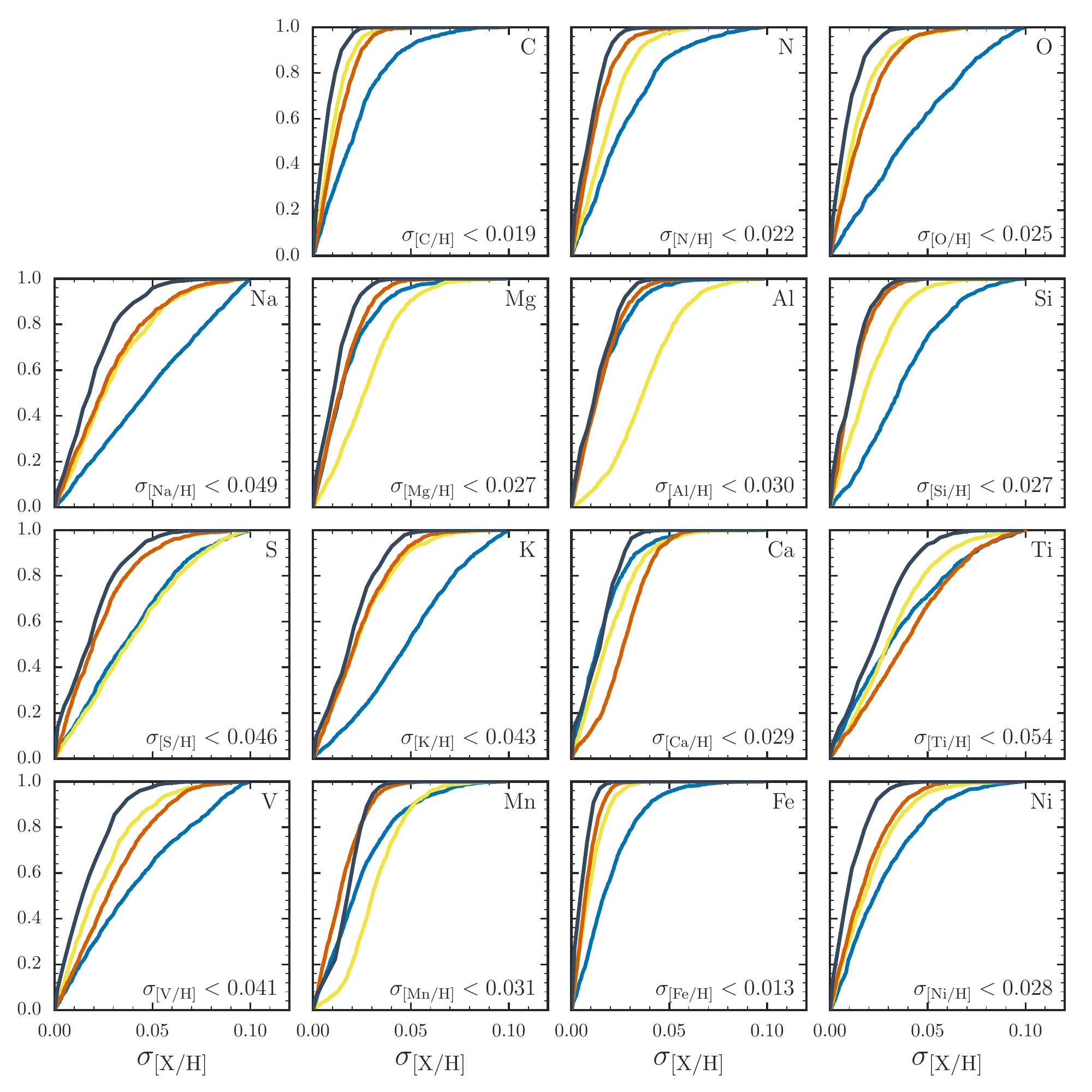}
  \end{center}
\caption{Cumulative posterior distribution functions for the intrinsic
  abundance scatter in 15 elements obtained from the ABC
  simulations. The color-coding of the three clusters is the same as
  in \figurename~\ref{fig:cumul}: M67 is in red, NGC 6819 is in
  yellow, and NGC 2420 is in blue. The black curve gives the
  cumulative distribution function from the combination of all three
  clusters; the 95\,\% upper limit from this combined PDF is given in
  each panel. The 68\,\% and 95\,\% upper limits for all three
  clusters and the combined PDF are given in
  \tablename~\ref{table:results}.}\label{fig:cumulres}
\end{figure*}

ABC produces an approximation to the PDF by (a) simulating
$\sigma_{\xh}$ from its prior (which I take to be uniform between $0$
and $0.1\dex$), and (b) only storing those $\sigma_{\xh}$ that lead to
simulated data that are ``the same'' as the real data
\citep{Tavare97a,Pritchard99a}. By considering the meaning of the PDF
(``the probability distribution of $\sigma_{\xh}$ given the data''),
it is clear that this procedure works, because it generates a Monte
Carlo sampling of $\sigma_{\xh}$ that are constrained to be the same
as the data and are therefore a sampling of the PDF. See, for example,
\citet{Marin12a} for a recent review of ABC.

ABC requires one to specify what it means for simulated data $D'$ to
be same as the actual data $D$. This is done by defining a metric
$\rho(D',D)$ that expresses how close the simulated data are to the
real data. The PDF for $\sigma_{\xh}$ is constructed using the
$\sigma_{\xh}$ that generate $D'$ that satisfy $\rho(D',D) \leq
\epsilon$ and ABC produces an exact sampling of the PDF in the limit
$\epsilon \rightarrow 0$. Of course, it is difficult to generate
simulated data that are exactly like the actual data, especially in
the presence of random noise. If the data and model have a
lower-dimensional sufficient statistic $\mu(D)$ that encapsulates all
of the information about $\sigma_{\xh}$ that is contained in the data,
this situation is significantly ameliorated. While a rigorous
sufficient statistic does not exist for the problem considered in this
paper, as I argue below there are summary statistics that can be used
to significantly reduce the dimensionality of the data and make this
problem tractable for ABC.

In principle, we need to constrain all 15 $\sigma_{\xh}$
simultaneously, because all 15 elements affect the spectra and
disentangling their effects is difficult, especially for C, N, and
O. Because I only derive upper limits on $\sigma_{\xh}$, however, I
can consider each element separately. That is, the lack of scatter in
the spectra near, \eg, CN features implies a limit on both
$\sigma_{[\mathrm{C/H}]}$ and $\sigma_{[\mathrm{N/H}]}$ that can be
established by varying $\sigma_{[\mathrm{C/H}]}$ and
$\sigma_{[\mathrm{N/H}]}$ separately. Stronger, covariant limits on
$\sigma_{[\mathrm{C/H}]}$ and $\sigma_{[\mathrm{N/H}]}$ could be
determined by considering them simultaneously, but I do not attempt
this here as it significantly increases the computational
complexity. Similarly, we need not worry too much about whether the
limit $\epsilon$ used to decide which simulated data are close to the
actual data is a good limit, because setting it too high will only
weaken the upper limits on $\sigma_{\xh}$.

I consider two summary statistics when running the ABC
simulations. The first is the Kolmogorov-Smirnov distance $D_n$
between the cumulative distribution of normalized residuals of the
data and the simulated data, that is, the maximum difference between
these distributions. For each element, these cumulative distributions
are computed by weighting the residuals by the weights for that
element (see \appendixname~\ref{sec:windows}). For the data these
cumulative distributions are shown in
\figurename~\ref{fig:cumul}. This figure also displays the median and
interquartile range of the cumulative distributions of simulated data
computed for $\sigma_{\xh} = 0.1\dex$.

The second summary statistic that I employ is based on the covariance
matrix of the normalized residuals of different pixels. For the data
and the simulated data, I compute the covariance matrix
$\mathrm{Cov}_{ij}$ between pixels $i$ and $j$. For each simulated
data set, I calculate the difference
\begin{equation}
  |\Delta \mathrm{Cov}_{ij}| = \sqrt{\textstyle{\sum_{ij}} (w_i\,w_j)^{1/2}\,(\mathrm{Cov}^{\mathrm{data}}_{ij}-\mathrm{Cov}^{\mathrm{sim.\ data}}_{ij})^2}\,,
\end{equation}
where $w_i$ and $w_j$ are the weights for a given element. The
covariance matrix $\mathrm{Cov}_{ij}$ for the data in M67 and for the
two simulated data sets from \figurename~\ref{fig:simpixal} are
displayed in \figurename~\ref{fig:simpixal_cov}. It is clear that
$|\Delta \mathrm{Cov}_{ij}|$ is a good sufficient statistic,
especially for elements with many absorption features, because any
abundance scatter in a given element will give rise to correlated
residuals at the positions of absorption features of that element. For
example, in \figurename~\ref{fig:simpixal_cov}, the simulated data
with $\sigma_{[\mathrm{Al/H}]} = 0.1\dex$ has both large scatter at
the positions of the Al lines, \emph{and} large off-diagonal
correlations between pixels in different lines. The fact that these
are absent for the data puts a strong constraint on the Al scatter in
M67.

As discussed in \sectionname~\ref{sec:empmodel}, variations in the LSF
for different stars can give rise to scatter in the residuals. While
the mock-data simulations take any LSF variations for APOGEE into
account, neither of the two summary statistics corrects for the effect
of LSF variations. Doing so would require a metric function
$\rho(D',D)$ that distinguishes between stars with different
LSFs. Alternatively, a procedure to homogenize the LSF could be
applied to both the data and the simulated data (in its crudest form,
this would consist of convolving all data to the worst LSF). The
effect of LSF variations can be seen in the zero-scatter simulation in
the middle panel of \figurename~\ref{fig:simpixal_cov} as the regions
of negative correlation surrounding the line centers and the positive
correlation between the central pixels of the two Al lines
shown. These features are absent in simulations using the same LSF for
all stars.

The distribution of these two summary statistics in the ABC
simulations for M67 are shown in \figurename~\ref{fig:m67stat} for Al
and Fe. It is clear that $|\Delta \mathrm{Cov}_{ij}|$ is a good
summary statistic, especially at large intrinsic scatter, as it
strongly correlates with $\sigma_{\xh}$. The statistic $D_n$
distinguishes between different $\sigma_{\xh}$ to a lesser extent, but
is important for identifying those simulations that are most like the
data, as $D_n$ captures some of the information in the shape of the
distribution of residuals that is not captured by $|\Delta
\mathrm{Cov}_{ij}|$.

To construct the PDF for $\sigma_{\xh}$ for each element X and each
cluster, I then run ABC simulations as described above. A flowchart of
how these simulations are run is given in
\figurename~\ref{fig:flowchart}. Before the start of each simulation,
I compute a fine grid in $\xh$ from $-0.20\,\dex$ to $+0.20\dex$ with
respect to the standard cluster abundances (median \feh\ of all
cluster members with solar abundance ratios) with a spacing of
$0.01\dex$; this grid is computed for each star
individually. Subsequent simulations that require \xh\ within the grid
use linear interpolation to generate the spectra; spectra for
\xh\ outside of the grid are computed on-the-fly and added to the
grid; interpolation is only ever performed for \xh\ located between
grid points within $0.01\dex$ from the nearest grid points. Each
simulation then proceeds by drawing a $\sigma_{\xh}$ from the prior
(uniformly between $0$ and $0.1\dex$), generating simulated spectra,
performing the quadratic \teff\ fit for each pixel, and computing the
two summary statistics $D_n$ and $|\Delta \mathrm{Cov}_{ij}|$.

Good limits on $D_n$ and $|\Delta \mathrm{Cov}_{ij}|$ are determined
by inspecting their distribution by eye and I run ABC simulations
until the distributions appear to have converged (that is, small
changes in the limits on $D_n$ and $|\Delta \mathrm{Cov}_{ij}|$ stop
mattering) and until about 1,000 $\sigma_{\xh}$ samples from the PDF
have accrued. The limits on $D_n$ and $|\Delta \mathrm{Cov}_{ij}|$ for
Al and Fe in M67 are displayed as dashed lines in
\figurename~\ref{fig:m67stat}.

I have performed tests of the code verifying that no constraints on
the abundance scatter are possible in the following limiting cases:
(a) When only using three stars in a cluster, because then the
quadratic fit is always perfect; and (b) when calculating the $D_n$
and $|\Delta \mathrm{Cov}_{ij}|$ summary statistics using the weights
of a different element than the one whose intrinsic scatter is being
constrained (choosing two elements with no overlapping weights, like
Al and C or Ca and Mn), because then the statistics are not sensitive
to abundance variations. All of such tests passed.

The cumulative distributions of the PDF for each element and each
cluster are displayed in \figurename~\ref{fig:cumulres}. The 68\,\%
and 95\,\% upper limits for each individual cluster are given in
\tablename~\ref{table:results}. The cumulative PDFs in
\figurename~\ref{fig:cumulres} demonstrate that we obtain strong
limits on the abundance scatter, especially in M67 and NGC 6819, where
we have the most cluster members. All elements are consistent with
having no scatter; only for Ca in M67 and Al and Mn in NGC 6819 do the
cumulative PDFs have a mild preference for a scatter of
$\approx0.03\dex$, but with a tail toward zero scatter. Multiple
elements for NGC 2420 have an almost flat PDF, especially O, Na, and
K; this is simply due to the fact that these features are very weak in
this more metal-poor cluster and that only a single low-\teff\ star is
included (for which we can get a precise O abundance).

\begin{deluxetable}{lcccccccc}
  \tablecaption{Limits on intrinsic abundance scatter}
  \tablecolumns{7}
  \tablewidth{0pt}
  \tablehead{\colhead{} & \multicolumn{2}{c}{M67} & \multicolumn{2}{c}{NGC 6819} & \multicolumn{2}{c}{NGC 2420} & \multicolumn{2}{c}{Combined}\\
  \colhead{} & \colhead{68\%} & \colhead{95\%} & \colhead{68\%} & \colhead{95\%} & \colhead{68\%} & \colhead{95\%} & \colhead{68\%} & \colhead{95\%}}
  \startdata
  C  & 0.016 & 0.030 & 0.013 & 0.025 & 0.027 & 0.058 & 0.009 & 0.019\\
N  & 0.015 & 0.031 & 0.023 & 0.043 & 0.034 & 0.069 & 0.013 & 0.022\\
O  & 0.022 & 0.041 & 0.017 & 0.039 & 0.055 & 0.088 & 0.010 & 0.025\\
Na  & 0.035 & 0.069 & 0.037 & 0.070 & 0.065 & 0.094 & 0.025 & 0.049\\
Mg  & 0.019 & 0.036 & 0.034 & 0.059 & 0.021 & 0.045 & 0.014 & 0.027\\
Al  & 0.019 & 0.035 & 0.045 & 0.069 & 0.020 & 0.041 & 0.018 & 0.030\\
Si  & 0.015 & 0.030 & 0.026 & 0.048 & 0.045 & 0.077 & 0.014 & 0.027\\
S  & 0.028 & 0.058 & 0.052 & 0.085 & 0.050 & 0.084 & 0.024 & 0.046\\
K  & 0.030 & 0.053 & 0.030 & 0.059 & 0.061 & 0.091 & 0.025 & 0.043\\
Ca  & 0.033 & 0.049 & 0.026 & 0.047 & 0.020 & 0.042 & 0.019 & 0.029\\
Ti  & 0.051 & 0.083 & 0.039 & 0.070 & 0.046 & 0.084 & 0.031 & 0.054\\
V  & 0.038 & 0.066 & 0.030 & 0.061 & 0.053 & 0.089 & 0.022 & 0.041\\
Mn  & 0.019 & 0.034 & 0.038 & 0.059 & 0.030 & 0.063 & 0.021 & 0.031\\
Fe  & 0.010 & 0.019 & 0.012 & 0.025 & 0.024 & 0.048 & 0.007 & 0.013\\
Ni  & 0.022 & 0.045 & 0.025 & 0.049 & 0.035 & 0.068 & 0.014 & 0.028\\

  \enddata
  \tablecomments{The 68\,\% and 95\,\% upper limits on the intrinsic
    abundance scatter in 15 elements. Those obtained from each
    individual cluster are given as well as those from combining all
    clusters.}\label{table:results}
\end{deluxetable}

From the results in \tablename~\ref{table:results}, we see that for
M67 and NGC 6819 we obtain strong constraints on the scatter in C, N,
O, Mg, Al, Si, Mn, Fe, and Ni. We obtain weaker constraints for Na, S,
K, Ca, Ti, and V and also obtain weaker results overall for NGC
2420. Given that I find no evidence for any abundance scatter in any
of the clusters, it is reasonable to combine the PDFs into joint
constraints on the abundance scatter (assuming all clusters have the
same intrinsic scatter and placing a limit on this). The PDF obtained
from combining the PDFs of all three clusters and the 95\,\% upper
limits on $\sigma_{\xh}$ from this combined PDF are indicated in
\figurename~\ref{fig:cumulres}; the combined constraints are also
included in \tablename~\ref{table:results}. We see that the combined
constraints are strong. With the exception of Ti, all elements have a
scatter constrained to be less than $0.05\dex$ at 95\,\% confidence.

The constraints on Fe and C are particularly strong: any intrinsic
scatter has to be $<0.007\dex$ and $<0.009\dex$ in Fe and C,
respectively, at 68\,\% confidence ($<0.013\dex$ and $<0.019\dex$ at
95\,\% confidence). We also obtain strong constraints on the intrinsic
scatter in N and O, which has to be $<0.013\dex$ and $<0.010\dex$,
respectively, at 68\,\% confidence ($<0.022\dex$ and $<0.025\dex$ at
95\,\% confidence). That I find the strongest limits on C, N, O, and
Fe is not surprising, because these elements have by far the most
abundant absorption features in the near-infrared wavelength region
used here, but the fact that I have been able to extract these
constraints from the complicated molecular features for C, N, and O
and in the light of deep mixing along the giant branch, demonstrates
the power of the method developed here.

I also obtain strong limits on the scatter in Mg, Si, and Ni; these
are all roughly $\lesssim0.015\dex$ and $\lesssim0.03\dex$ at 68\,\%
and 95\,\% confidence, respectively. The limits on the scatter in Al,
Ca, and Mn are less strong, but are nevertheless $\lesssim0.02\dex$
and $\lesssim0.03\dex$ at 68\,\% and 95\,\% confidence. The weaker
features of Na, S, K, Ti, and V give weaker limits that are about
$0.025$ to $0.03\dex$ at 68\,\% and about $0.05\dex$ at 95\,\%
confidence.

\section{Discussion and conclusions}\label{sec:conclusions}

\subsection{Prospects for chemical tagging}

The novel technique introduced here for constraining the abundance
scatter in open clusters has several advantages over traditional
techniques that determine each individual star's abundances and
constrain the scatter in these abundances. Firstly, the new technique
is robust to systematic uncertainties in the abundances stemming from
the fact that obtaining consistent abundances over wide ranges of
stellar types is difficult. The systematics introduced by, \eg,
deviations from the assumptions of local thermodynamic equilibrium,
one-dimensional radiative transfer, and uncertainties in the line list
will cause offsets in the abundances that are smooth functions of
stellar mass (or here specifically, \teff). By remaining agnostic
about the overall \teff\ trends of the spectra of the cluster members
analyzed here, I avoid all of these issues directly. Secondly, many of
the stellar-evolutionary effects on the surface abundances due to,
\eg, deep mixing or gravitational settling, also change the abundances
in a manner that is perhaps not well understood, but that is largely a
deterministic function of the stellar mass. Therefore, I was able to
constrain the \emph{initial} abundance scatter rather than the
present-day scatter and in particular constrain the scatter in C and
N. Thirdly, the use of forward simulations and ABC makes it
straightforward to include a large variety of real-world complications
in the observed spectra such as non-Gaussian and variable LSFs,
correlated noise in the spectra, and uncertainties coming from the
applied continuum normalization.

It is easy to think of factors that should lead to a break-down of the
one-dimensional assumption that I propose here for clusters. The large
spread in initial rotation velocities will give rise to differences in
the spectra of young-cluster members that are largely orthogonal to
those from differences in the initial mass. While the initial
differences in rotation speeds will have largely disappeared due to
magnetic braking for older open clusters \citep{Kawaler88a}, the
effect of the different viewing angles would still give rise to
scatter in the spectra and for the technique used in this paper to
work, the $v\sin i$ of each star probably needs to be inferred prior
to the forward simulations. For the giants that I studied here,
rotational velocities are small and I was therefore able to ignore
this complication. Beyond the effects of the current velocity, initial
velocity differences may still lead to spectral scatter today if they
led to different mixing histories, changing the surface abundances of,
\eg, C, N, and Li \citep{Pinsonneault90a,Meynet02a}. Effects of
binarity and, in particular, of mass transfer between binary
companions could also induce spectral scatter which would be difficult
to correct.

The fact that I empirically determine that all open clusters are
consistent with being one-dimensional sequences and that I am able to
determine tight limits on the abundance scatter in 15 elements,
implies that the effects discussed in the previous paragraph must be
small, at least for the red giants investigated here. This places
strong limits on the probability and efficacy of the effects of
rotation and binarity. From an operational viewpoint, this is good
news for the prospect of performing detailed Galactic archaeology
through large-scale chemical tagging. The old open clusters that we
observe today most likely represent a population of stars where the
effects of rotation and binarity would be the strongest, such that
variations due to these confounding factors should be even smaller
among field stars born in more loosely-bound associations. These
variations would furthermore likely be largest for the surface
abundances of C, N, and O (because they have large abundances, are
strongly affected by mixing, and would be transferred by stellar winds
to binary companions). These are the elements for which I obtain the
strongest constraints, with intrinsic scatter constrained to be
$<0.025\dex$ at 95\,\% confidence in each of these. It does therefore
not require a large leap of faith to assume that such limits are
possible for all elements, especially in the light of the
considerations in the next section. Thus, we should be able to
determine the fine-grained structure of (initial) abundance space of
the Milky Way, given a large enough sample of stars with
high-resolution spectra.

Further development of the empirical technique presented in this paper
should also allow chemical tagging to be performed in a manner that is
less prone to systematics and that requires less input from stellar
physics. That is, a promising way forward for chemical tagging is to
search for one-dimensional sequences among a library of spectra,
rather than searching for zero-dimensional loci in abundance
space. Aside from being more robust against systematics, this will
also allow C and N to be fully used for giants, because they are
currently often excluded because of the effects of deep mixing
\citep[\eg,][]{Ting15a}. While a second-order-polynomial model
suffices to describe the empirical stellar-mass trends of the
red-giant spectra in the current sample, data sets that contain a
wider variety of stellar types (including, for example, main-sequence
stars and sub-giants) likely require more flexible one-dimensional
models. This is especially the case when sharp changes in the surface
abundances, \eg, due to the first dredge up \citep{Iben64a}, occur
within the sample.

\subsection{The formation of star clusters}

The tight constraints on the initial abundance scatter in open
clusters places strong limits on how star formation in molecular
clouds---the progenitors of open clusters---proceeds. This is
especially the case because of the strong limits on the scatter in C,
N, and O. C and O are produced in large quantities in core-collapse
supernovae (CCSNe). Using the yields at solar metallicity from
\citet{Chieffi04a} and \citet{Limongi06a}, we find that
$\approx0.9\msun$ and $3.5\msun$ of C is produced in a single CCSN of
a $35\msun$ and a $60\msun$ star, respectively; for O the yield is
even higher: $\approx5\msun$ and $8.5\msun$ for the same two
masses. Assuming that this amount of C or O is mixed in with
$\approx20,000\msun$ of gas with solar abundance ratios from
\citet{Asplund05a}, subsequent stars would have C abundances higher by
$\approx0.02\dex$ and $0.04\dex$ and O abundances higher by
$\approx0.02\dex$ and $0.03\dex$ for a single CCSN of a $35\msun$ and
$60\msun$ star. For comparison, the amount of Fe produced in these
CCSNe is only about $0.1$ to $0.2\msun$ and raises the Fe abundances
of new stars by $\approx0.004\dex$.

The fact that the initial scatter in C and O is constrained to be
$\lesssim0.025\dex$ at 95\,\% confidence implies that no pollution by
massive CCSNe occurred before most of the stars formed. The initial
masses of M67 and NGC 6819 were likely in the range $10,000\msun$ to
$20,000\msun$ \citep{Hurley05a,Yang13a}. Using the IMF from either
\citet{Kroupa01a} or \citet{Chabrier03a}, we would expect $\approx11$
and $5$ stars with masses greater than $35\msun$ and $60\msun$,
respectively, in a cluster with a mass of $20,000\msun$. Thus, we
would expect CCSNe of $35\msun$ to $60\msun$ stars to occur in M67 and
NGC 6819 and potentially even higher mass CCSNe, which would lead to
even larger abundance scatter. For star formation lasting for a time
$\tau_{\mathrm{SF}}$, the lack of a CCSN when $K$ massive stars with
lifetimes $\tau_{\mathrm{CCSN}}$ are expected to form gives the
following PDF for $\tau_{\mathrm{SF}}$ (assuming a flat prior)
\begin{align*}
  p(\tau_{\mathrm{SF}}|\mathrm{no\ CCSN}) \propto \begin{cases}
      1 & \text{if $\tau_{\mathrm{SF}} < \tau_{\mathrm{CCSN}}$} \\
      \left(\frac{\tau_{\mathrm{CCSN}}}{\tau_{\mathrm{SF}}}\right)^K & \text{otherwise}\,.
    \end{cases}
\end{align*}
This converges to a flat distribution between zero and
$\tau_{\mathrm{CCSN}}$ for large $K$, because of the increasing
probability that a massive star is formed at the onset of star
formation and because we cannot distinguish $\tau_{\mathrm{SF}}$ that
are smaller than $\tau_{\mathrm{CCSN}}$. For 11 expected $35\msun$
stars (with lifetimes $\tau_{\mathrm{CCSN}}\approx5.7\Myr$) as well as
for 5 expected $60\msun$ stars ($\tau_{\mathrm{CCSN}}\approx4\Myr$;
\citealt{Bressan12a}), this gives an upper limit on
$\tau_{\mathrm{SF}}$ of $\approx6\Myr$ at 95\,\% confidence. This
limit would obviously weaken if massive stars preferentially form
after low-mass stars---although this is not expected to be the case
\citep{McKee02a} and they may even form the earliest
\citep{Maschberger10a}---or if a significant portion of the SNe ejecta
are introduced into a warm ISM phase that is not immediately available
for star formation (\eg, \citealt{Matzner00a}, but see
\citealt{Pan12a}).

The fact that I find no scatter in the abundances of light elements
also directly demonstrates that the type of pollution that occurs in
globular clusters does not happen for open clusters. Globular clusters
display significant abundance scatter and anti-correlations in the
abundances of light elements (C, N, O, Na, Mg, Al;
\citealt{Gratton04a}), believed to stem from pollution of the
intracluster medium by intermediate-mass asymptotic giant branch stars
\citep[\eg,][]{Ventura01a}, fast-rotating massive stars
\citep{Decressin07a}, or massive binaries \citep{deMink09a}. The
abundance scatter in all of the light elements commonly studied in
globular clusters is $<0.03\dex$ at 95\,\% confidence in the open
clusters analyzed here (except for Na, for which the limit is slightly
weaker).

That the initial abundance scatter in open clusters is as small as
$0.01$ to $0.02\dex$ as found here challenges our understanding of the
structure of molecular clouds. To attain this level of homogeneity,
the gas and dust in star-forming clouds has to be very well mixed
\citep{Feng14a,Hopkins15a} and, as argued above, star formation has to
proceed within about $6\Myr$. This is an important new constraint on
the timescale of star formation in molecular clouds
\citep[\eg,][]{Elmegreen00a,Tan06a,Matzner07a}. The timescale
constraint derived here is limited not by the constraint on the
abundance spread, but instead by whether a CCSN of a massive star is
likely to have occurred and to have polluted the star-forming
gas. Therefore, the kind of limits derived here will not be able to be
improved much further.

\subsection{Final remarks}

For many of the questions relating to the formation and evolution of
star clusters and galactic disks that we may answer using detailed
measurements of stellar abundances, the \emph{precision} in the
abundances is of much higher importance than their overall
accuracy. However, much of the modeling effort currently going into
the analysis of large spectroscopic surveys is focused on improving
the theoretical modeling of stellar photospheres (\eg,
\citealt{Magic13a}) or line formation beyond the simplest models
\citep[\eg,][]{Bergemann12a}. While more realistic modeling of the
stellar photospheres, radiative transfer, and line formation will
provide a significant improvement for any spectroscopic analysis, it
is unlikely that all systematic effects in the abundances will be
removed through these efforts in the near future. This is especially
the case for the effects of, \eg, deep mixing or atomic diffusion that
actually change the surface abundances in a manner that is not
entirely well understood \citep[\eg,][]{Onehag14a}. 

The method introduced here is wholly focused on obtaining the highest
possible abundance precision given the observational limitations. It
does this at the expense of some of the information in the spectra,
which instead of being used to constrain the abundance scatter, is
used to build an empirical model of the spectra. That the new
technique leads to some of the tightest constraints on the intrinsic
abundance scatter in open clusters and that it does this based on the
complex infrared APOGEE spectra of giants that likely have intrinsic
variations in C and N, is a testament to the strength of this new
technique. I expect that extensions of this technique to other
groupings of stars and to the whole Galactic disk population will lead
to fundamentally new insights into the formation and evolution of
stellar populations in the Milky Way.

\acknowledgements It is with great pleasure that I thank the APOGEE
ASPCAP team for many valuable discussions regarding infrared
spectroscopy and APOGEE. In particular, I thank Carlos Allende Prieto,
Katia Cunha, Anibal Garc\'{\i}a-Her\'{n}andez, Jon Holtzman, Szabolcs
M{\'e}sz{\'a}ros, Matthew Shetrone, and Olga Zamora for help with the
APOGEE ATLAS9 model atmospheres, \texttt{Turbospectrum} and
\texttt{MOOG}, the APOGEE line list, the APOGEE LSF, and the APOGEE
pipeline. I further acknowledge insightful conversations with and
comments from the anonymous referee, Ewan Cameron, David Hogg, Chris
McKee, Melissa Ness, Garrett Somers, and Yuan-Sen Ting. I also thank
the Natural Sciences and Engineering Research Council of Canada for
financial support of this project.

Funding for SDSS-III has been provided by the Alfred P. Sloan
Foundation, the Participating Institutions, the National Science
Foundation, and the U.S. Department of Energy Office of Science. The
SDSS-III web site is http://www.sdss3.org/.

\appendix

\section{High signal-to-noise ratio mean $H$-band spectra for M67, NGC 6819, and NGC 2420}\label{sec:clusterspec}

\begin{figure*}[t!]
\begin{center}
\includegraphics[width=0.88\textwidth,clip=]{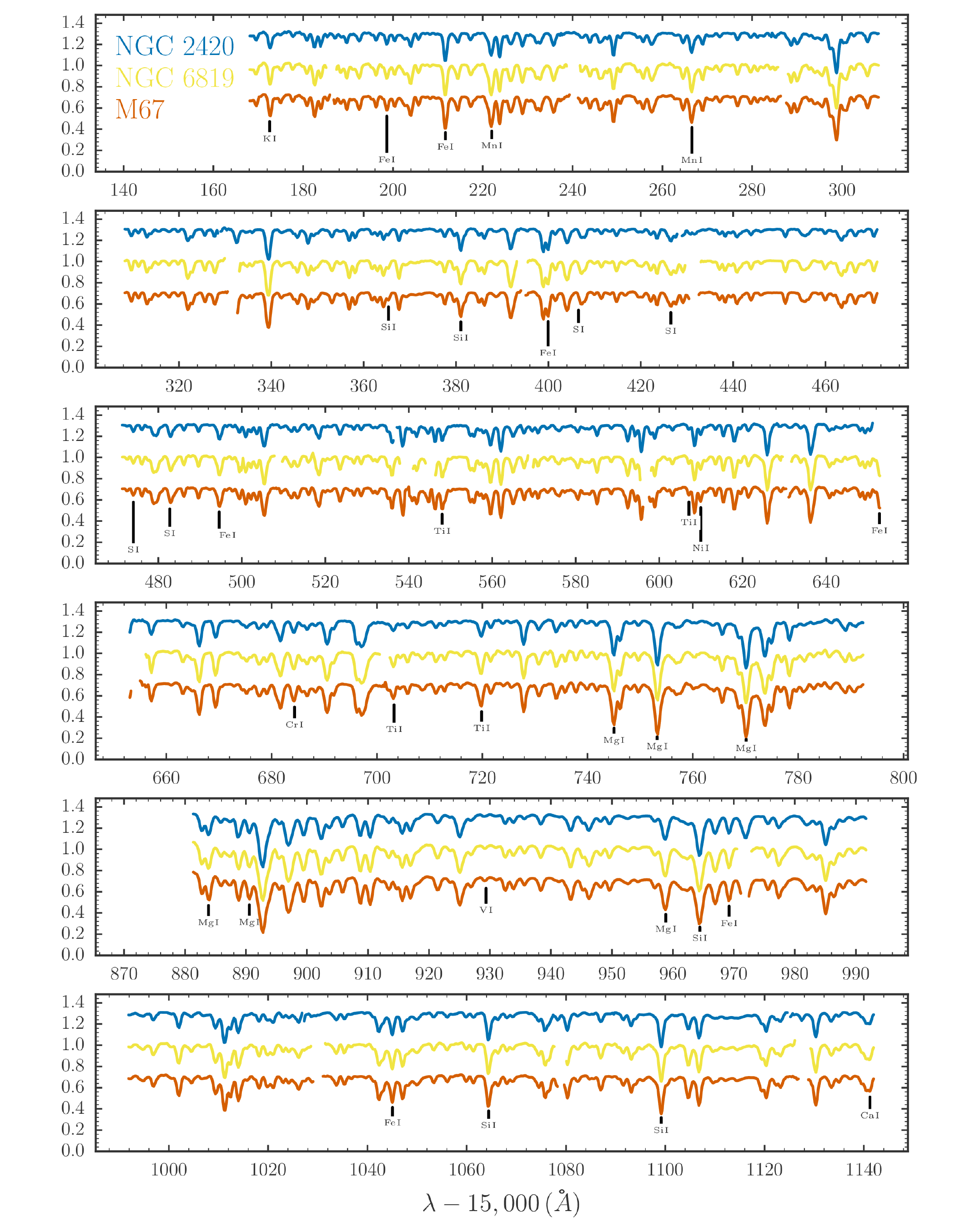}
\end{center}
\caption{Combined continuum-normalized spectra between $\lambda
  \approx 15,150\,\AA$ and $16,150\,\AA$ for a first-ascent red giant
  at $\teff = 4750\,\mathrm{K}$ in M67, NGC 6819, and NGC 2420. These
  are determined from the quadratic fit to the \teff\ dependence of
  each pixel using the red-giant members in each cluster (excluding
  red-clump stars). Strong, clean atomic lines from the compilation of
  \citet{Smith13a} for most of the elements considered in this paper
  (and some others) are indicated. The spectra for M67 and NGC 2420
  are offset by $-0.3$ and $0.3$, respectively. Almost every feature,
  including weak ones, are reproduced in all three spectra,
  demonstrating that these spectra contain very little
  noise.}\label{fig:clusterspec0}
\end{figure*}
\begin{figure*}[t!]
\begin{center}
\includegraphics[width=0.88\textwidth,clip=]{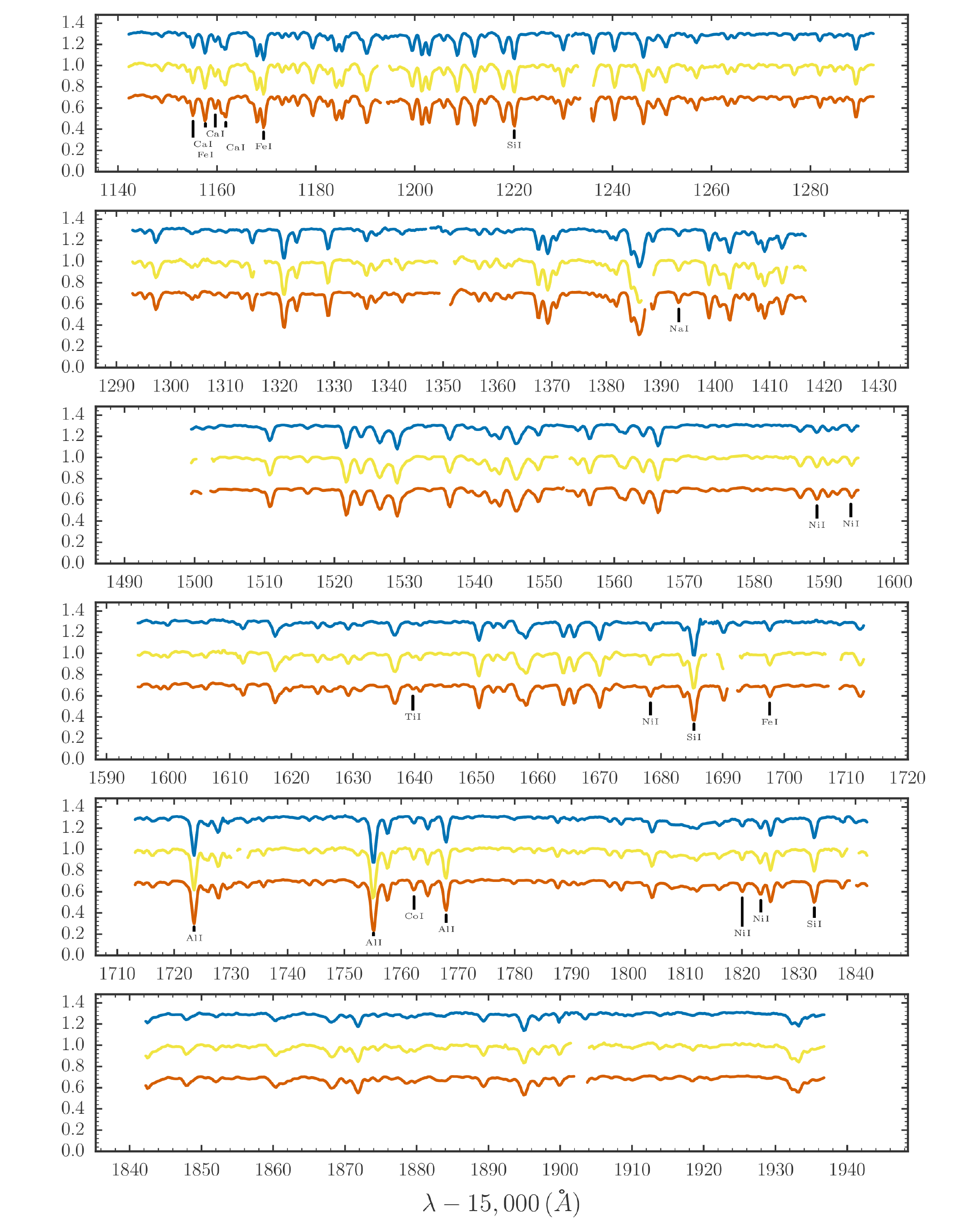}
\end{center}
\caption{Same as \figurename~\ref{fig:clusterspec0}, but for the
  wavelength range $\lambda \approx 16,150\,\AA$ to
  $16,950\,\AA$.}\label{fig:clusterspec1}
\end{figure*}

In this appendix, I display the combined cluster spectra for M67, NGC
6819, and NGC 2420. These are obtained from the quadratic fits to each
pixel using the modeling of \sectionname~\ref{sec:empmodel}. Because
the spectra of members of these clusters are all consistent with being
a function a function of \teff\ only, we can use the quadratic fit to
their \teff\ dependence to construct a very high signal-to-noise-ratio
spectrum for each cluster. I compute the cluster spectrum at $\teff =
4750\,\mathrm{K}$, which is close to the median \teff\ of all of the
considered cluster members. To avoid any confusion due to the
inclusion of red-clump stars in M67 and NGC 6819, I only include
first-ascent red giants in the quadratic fit and the cluster spectrum
is therefore that of a red giant in these clusters at $\teff =
4750\,\mathrm{K}$.

The cluster spectra are shown at high resolution in \figurename s
\ref{fig:clusterspec0} and \ref{fig:clusterspec1}. The three spectra
are clearly very similar, especially those for M67 and NGC 6819 which
are close in age and metallicity. Almost every single wiggle in the
spectra is repeated in all three spectra, demonstrating the extremely
high signal-to-noise ratio of these combined spectra. They could be
used to obtain precise abundances for these clusters, but this is not
done here as it is beyond the scope of this paper.

\section{Empirical investigation of the APOGEE spectral errors using repeat observations}\label{sec:repeats}

\begin{figure*}
  \begin{center}
  \includegraphics[width=0.9\textwidth,clip=]{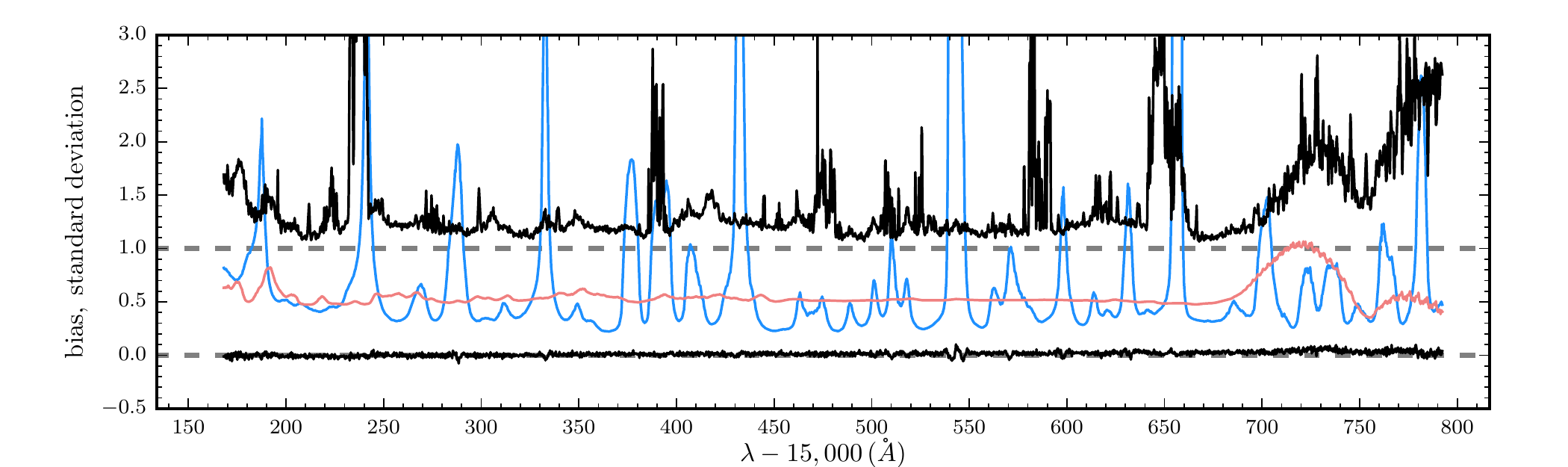}\\
  \includegraphics[width=0.9\textwidth,clip=]{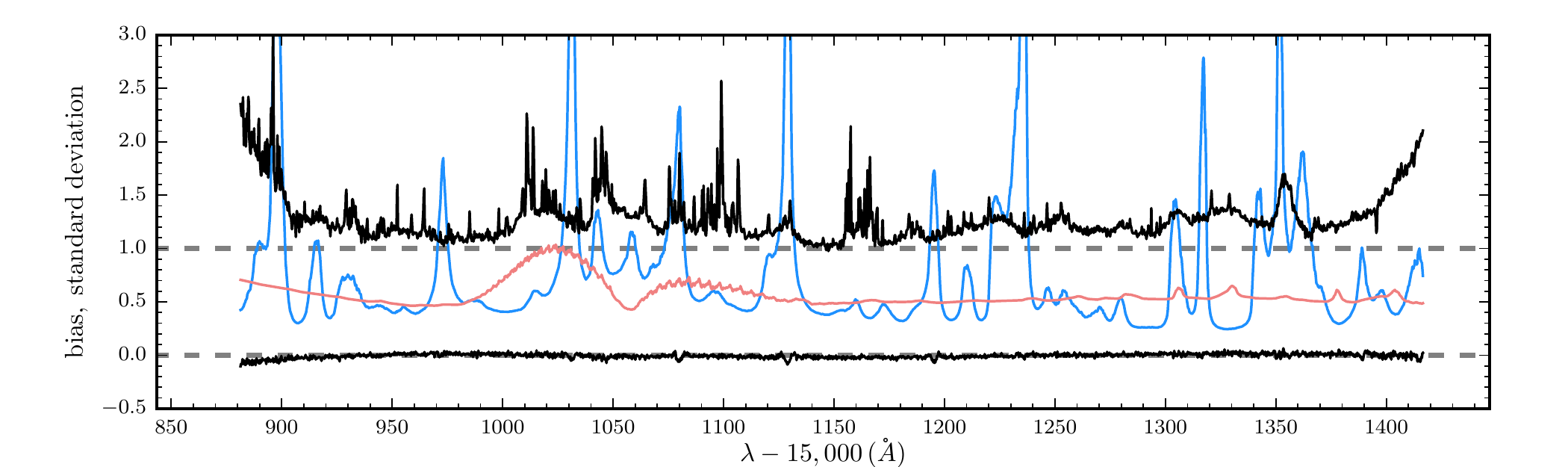}\\
  \includegraphics[width=0.9\textwidth,clip=]{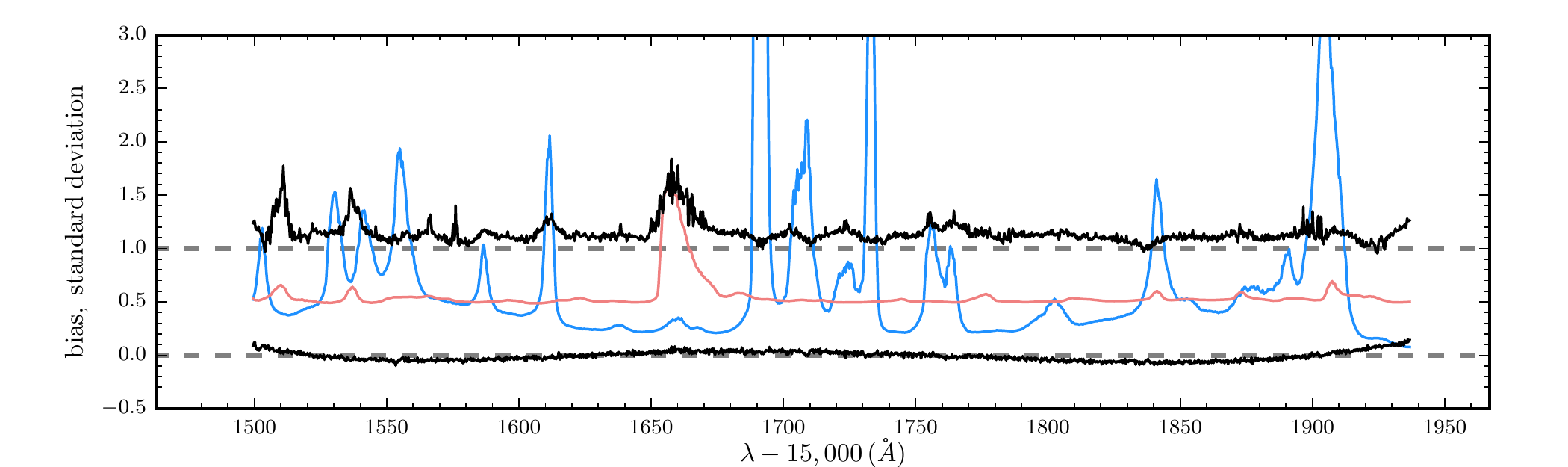}
  \end{center}
\caption{The median (bias; black line near zero) and standard
  deviation (black line above one) of the normalized residuals of
  individual exposures vs. combined continuum-normalized spectra of
  1,381 stars with three exposures that individually have high
  signal-to-noise ratio. Each panel displays one of the three APOGEE
  detectors, which span three different wavelength ranges. The median
  is near zero, demonstrating that the continuum-normalized spectra
  are unbiased, with only minor effects due to the polynomial
  continuum-normalization. The standard deviation should be $<1$ if
  the reported APOGEE spectral uncertainties were correct (see text),
  but is typically 1.1 to 1.2, with large wavelength regions where it
  is $>1.5$. The blue and red lines display the average sky and
  telluric spectra used to correct the individual spectra for the
  effect of sky-emission and telluric-absorption lines. Regions of
  high scatter in the residuals appear to largely coincide with those
  with significant telluric absorption.}\label{fig:bias}
\end{figure*}

\begin{figure}
  \includegraphics[width=0.48\textwidth,clip=]{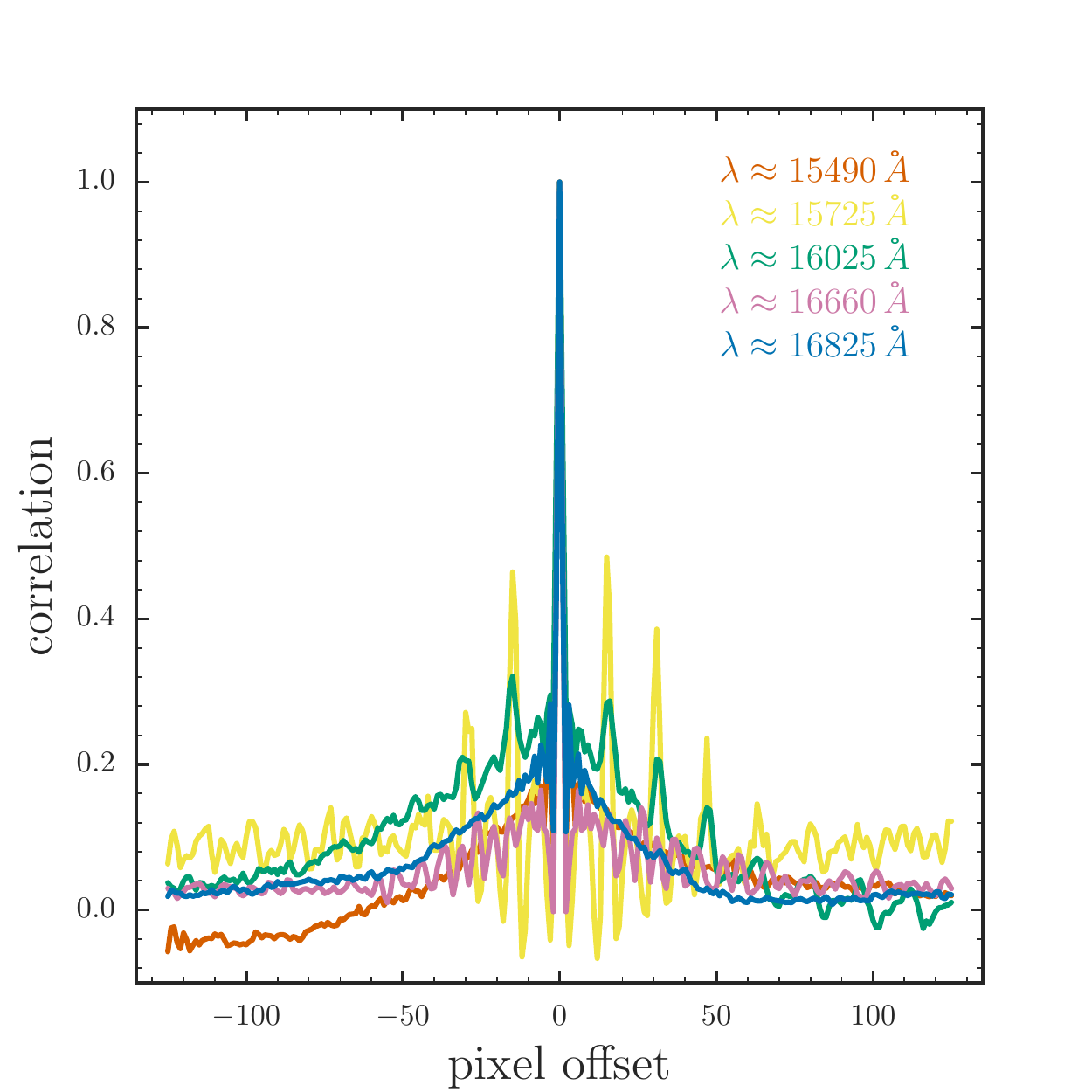}
\caption{Correlations between the errors in neighboring pixels
  determined from repeat observations (see \figurename~\ref{fig:bias})
  for five different central pixels. Significant correlations exist
  out to dozens of pixel offsets, both in regions of high and low
  scatter in the repeats (see \figurename~\ref{fig:bias}). This is
  much wider than the line-spread function and is most likely due to
  the continuum normalization.}\label{fig:corr}
\end{figure}

As discussed in \sectionname~\ref{sec:data}, the APOGEE spectra have
associated pixel-level uncertainties in the standard DR12 data
products. These are obtained from a noise model that tracks the
Poisson photon-counting noise through the APOGEE pipeline. These
uncertainties are for single pixels only; correlations between the
errors of neighboring pixels are not tracked. \citet{Nidever15a}
tested these uncertainties using the scatter in the repeat
observations of stars with six individual hour-long exposures,
demonstrating that the uncertainties overall track the scatter well,
but also finding a systematic error floor at the $0.5\,\%$ level (see
discussion in \sectionname~\ref{sec:data}). This test did not
distinguish between pixels at different wavelengths and regions with
significant sky-emission or telluric-absorption lines were avoided.

I perform a similar, but more detailed, test here and furthermore
determine an empirical model for the spectral errors using repeat
observations. I select giants from the APOGEE DR12 data set with
$4000\,\mathrm{K} \leq \teff \leq 5000\,\mathrm{K}$ and $\logg <
3.5$. Of these giants, I consider those with $10 < H < 11$ with three
hour-long exposures that on average have an overall signal-to-noise
ratio per half-resolution element larger than 100 (thus, their
combined spectrum has signal-to-noise ratio larger than 300). These
characteristics are similar to the majority of the open-cluster
members considered in this paper. This sample consists of 1,381 stars
with 3 repeat observations each for a total of 4,143 individual
spectra. All of the individual spectra have an overall signal-to-noise
ratio larger than 80.

I then continuum-normalize each individual-exposure spectrum as well
as the combined APOGEE spectrum for each of these stars in the manner
described in \sectionname~\ref{sec:data} and compute the normalized
residuals
\begin{equation}
\Delta f_\lambda^i/\delta_\lambda^i = \frac{f_\lambda^i-f_\lambda^{i,\mathrm{combined}}}{\delta_\lambda^i}\,
\end{equation}
for each star $i$, where $\delta_\lambda^i$ is the pipeline
uncertainty for each pixel $\lambda$. Pixels with signal-to-noise
ratio less than 50 or with any of the bad pixel flags discussed in
\sectionname~\ref{sec:data} are removed from further consideration.

The median of the normalized residuals $\Delta
f_\lambda^i/\delta_\lambda^i$ is displayed in
\figurename~\ref{fig:bias} and it measures the bias in the spectra
taking into account the effects of continuum normalization. It is
clear that the bias is small for all wavelengths, although minor
polynomial trends especially in the green detector (middle panel) and
the red detector (bottom panel) remain; these are due to the
polynomial continuum fitting not being quite reproducible between
different observations of the same star. I have also performed the
same test using the standard APOGEE continuum-normalization method
\citep{GarciaPerez15a} and found significantly larger biases (up to
about 0.5) that could negatively affect parameter and abundance
determinations from these spectra. Similarly, I found larger biases
when using only a second-order polynomial as in \citet{Ness15a}.

The standard deviation of the normalized residuals is also shown in
\figurename~\ref{fig:bias} and is typically about 1.1 to 1.2, but with
large wavelength ranges where the standard deviation is larger than
$1.5$. Because the three individual exposures are compared to their
combined value, the distribution of the sum of the squares of the
normalized residuals should follow a $\chi^2$ distribution with two
degrees of freedom if the pipeline uncertainties are correct and the
standard deviation of all residuals should be approximately
$2/3$. Therefore, the fact that the standard deviation is larger than
$2/3$ demonstrates that the pipeline uncertainties are
underestimated. \figurename~\ref{fig:bias} also contains the median
sky and telluric spectra that were used to correct the individual
exposures used here. As expected because I remove any pixels near sky
lines, the location of sky lines does not appear to be correlated with
large values of the standard deviation of the residuals. However,
these large values do appear to coincide with regions with significant
telluric absorption. It is therefore likely that the underestimated
uncertainties are due to issues with the telluric correction.

We can use the same normalized residuals to investigate the
correlations between the errors of neighboring
pixels. \figurename~\ref{fig:corr} displays the correlation for five
pixels chosen to represent a range of detectors and of regions with
low and high scatter in the normalized residuals. This figure clearly
demonstrates that there typically are significant correlations out to
dozens of pixel offsets, corresponding to $\gtrsim10\,\AA$. This is
much wider than the wavelength region over which the line spread
function is significant and these correlated errors are most likely
due to correlations induced by the continuum normalization, although
they may also have some contribution from scattered light.

When relating the differences in the spectra of stars to those
expected from scatter in the abundances, it is essential to have a
good understanding of the random errors and their correlations that
affect these differences. Rather than using the repeat observations to
build an empirical noise model incorporating the correlations between
pixels, I directly use the normalized residuals determined from the
repeat observations as an empirical sampling of the noise. This has
the advantage of being incredibly straightforward. When looking at the
residuals of the spectra in a given cluster from the one-dimensional
model using the method of \sectionname~\ref{sec:empmodel} as applied
in \sectionname~\ref{sec:oned}, I simply compare to the distribution
determined from the repeat observations. Similarly, in the forward
simulations described in \sectionname~\ref{sec:inference} and in
\appendixname~\ref{sec:synspec}, I simply draw from the set of
normalized residuals in the process of determining a mock
error. However, because the residuals come from a comparison between
an individual exposure and a combined spectrum that contains
information from this individual exposure, the residuals are slightly
smaller than the true error (because the three residuals for a given
star only have two degrees of freedom). For the purposes of this
paper, this is a conservative mistake, because it means that we are
slightly underestimating the uncertainties in the spectra. This will
slightly inflate any limit on the abundance scatter of the open
clusters studied here. This could be fixed in the future by using the
residuals determined from repeat observations to infer a noise model
or it could be ameliorated by increasing the number of repeat
exposures to $\gg3$. Another problem is that if the underestimation of
the pipeline uncertainties is truly due to the telluric-absorption
correction, a model for the uncertainties in the Earth's restframe
rather than in the star's restframe is necessary.

\section{Synthetic APOGEE spectra}\label{sec:synspec}

In \sectionname\sectionname~\ref{sec:oned} and \ref{sec:inference}, I
employ synthetic APOGEE spectra varying the abundances of 15 elements
with absorption lines in the APOGEE wavelength range. This
\appendixname\ explains how I generate these synthetic spectra. For
each individual star, I generate a model atmosphere at the median
metallicity of the cluster and at the $(\teff,\logg)$ of the star,
using solar abundance ratios for all elements. The model atmosphere is
obtained using linear interpolation of the grid of atmospheres
computed using the ATLAS9 code by \citet{Meszaros12a}. These
atmospheres and the synthetic spectra computed using them all use the
solar abundances from \citet{Asplund05a}. I compute synthetic spectra
varying the abundances of individual elements using
\texttt{Turbospectrum} \citep{Alvarez98a}, adopting the
microturbulence prescription as a function of \logg\ used in APOGEE
DR12, an isotopic ratio $C^{12}/C^{13} = 15$ appropriate for giants,
and using a Gaussian macroturbulence with a full-width-at-half-maximum
of $6\kms$. This macroturbulence is at the high end of what is
expected for the giants in this sample, but oversmoothing the spectrum
is conservative in that it would weaken any result on abundance
variations from spectral scatter. I employ the same line list as used
in APOGEE's DR12 \citep{Shetrone15a}, with astrophysical $gf$s
determined by fitting the $H$-band spectra of the Sun and Arcturus,
but fitting to the center-of-disk solar flux (M.~Shetrone, private
communication), \ie, with the total-flux vs. center-of-disk bug in the
APOGEE DR12 line list fixed (see \citealt{Shetrone15a}). Synthetic
spectra are computed in air wavelengths over the wavelength range
$15,000$ to $17,000\,\AA$ with a wavelength step of $0.05\,\AA$.

Each of the 300 APOGEE fibers has a different LSF. Variations in the
width of the LSF between different fibers are typically $10$ to
$20\,\%$, but the LSF of individual fibers are stable at the $1\,\%$
level \citep{Nidever15a}. The APOGEE LSF is non-Gaussian and detailed
forward modeling of the spectra needs to take the non-Gaussian,
variable nature of the LSF into account. I compute the LSF of each
fiber using the Gauss-Hermite-expansion fit to the LSF of each fiber
(which additionally includes a wide Gaussian for the wings of the LSF)
and the wavelength calibrations for all three detectors, which are
publicly available \citep{Nidever15a}. For each cluster star I average
the LSFs of the fibers used for the hour-long individual
exposures. The raw synthetic spectra from \texttt{Turbospectrum} are
interpolated onto a wavelength grid in vacuum wavelengths using the
transformations from \citet{Ciddor96a}. They are then convolved with
the LSF and brought onto the same wavelength grid as the observed
spectra. These spectra are then continuum-normalized using the
procedure described in \sectionname~\ref{sec:data}.

To add errors to the continuum-normalized synthetic spectra, I draw
from the set of 4,143 normalized residuals from repeat observations
(see \appendixname~\ref{sec:repeats}). This normalized error is then
``de-normalized'' by multiplying it with the uncertainty array of the
observed spectrum and the result is added to the synthetic
spectrum. Any bad pixels in the observed spectrum of a cluster star
are also labeled as bad in the synthetic spectra for that star. In
drawing from the normalized residuals, only those residuals with a
smaller number of bad pixels than the observed spectrum are used. In
generating synthetic spectra for changes in different elements, this
number of bad pixels is computed by weighting with the sensitivity
weights for each element that are discussed in
\appendixname~\ref{sec:windows}. That is, only bad pixels in the
spectral regions with absorption features for the given element are
taken into account. This procedure creates synthetic spectra that are
very similar to the observed spectra in LSF, errors, and distribution
of bad pixels.

All of the code to generate these synthetic spectra is available
online as part of a general-use APOGEE data-analysis Python package
called \texttt{apogee}, available
at\\ \centerline{\url{http://github.com/jobovy/apogee}\,.} This
package allows one to download and open the necessary data files
containing the APOGEE catalog and spectra. Tools for reading the
ATLAS9 APOGEE model atmosphere grid\footnote{Available here:
  \url{http://www.iac.es/proyecto/ATLAS-APOGEE/}.}, interpolating
within the grid using linear interpolation, and outputting the model
atmospheres in a format suitable for \texttt{MOOG} \citep{Sneden73a}
or \texttt{Turbospectrum} are included in
\texttt{apogee.modelatm}. Synthetic spectra can be calculated using
\texttt{MOOG} or \texttt{Turbospectrum} with a similar Python
interface using functions in \texttt{apogee.modelspec.moog} and
\texttt{apogee.modelspec.turbospec}. These include functions to simply
compute the high-resolution theoretical spectra or to generate full
synthetic spectra including LSF and macroturbulence convolution,
re-sampling onto the observed wavelength grid, and continuum
normalization.

Tools for computing the LSF and convolving with it efficiently using
sparse-matrix algebra are contained in
\texttt{apogee.spec.lsf}. Continuum-normalization using the standard
APOGEE method \citep{GarciaPerez15a} or the method of \citet{Ness15a}
(as used in this paper) are implemented in
\texttt{apogee.spec.cont}. Various tools for handling the sensitivity
windows described in \appendixname~\ref{sec:windows} are included in
\texttt{apogee.spec.windows}. The code to fit a linear or quadratic
model in stellar labels ($\teff$ in this paper, see
\sectionname~\ref{sec:empmodel} and \citealt{Ness15a}) is included in
\texttt{apogee.spec.cannon}. Aside from these tools used in the
analysis in this paper, the \texttt{apogee} package also contains a
full implementation of the standard APOGEE stellar-parameters and
elemental-abundances pipeline using
\texttt{FERRE}\footnote{\url{http://www.as.utexas.edu/~hebe/ferre/}\,.}. This
is included in \texttt{apogee.modelspec.ferre}, which allows for
interpolation of model spectra from the standard APOGEE grids
\citep{Zamora15a} and for performing the APOGEE stellar-parameter and
abundance fits for any APOGEE spectrum.

\section{Sensitivity of APOGEE spectra to abundance variations of different elements}\label{sec:windows}

\begin{deluxetable*}{lrrrrrrrrrrrrrrr}
  \tablecaption{Spectral variations for 0.1 dex abundance changes}
  \tablecolumns{16}
  \tablewidth{0pt}
  \tablehead{\colhead{Window} & \colhead{C} &
    \colhead{N} & \colhead{O} & 
    \colhead{Na} & \colhead{Mg} & 
    \colhead{Al} & \colhead{Si} & 
    \colhead{S} & \colhead{K} & 
    \colhead{Ca} & \colhead{Ti} & 
    \colhead{V} & \colhead{Mn} & 
    \colhead{Fe} & \colhead{Ni}}
  \startdata
  C & 0.022& 30& 31& 1& 17& 2& 8& 1& 9& 1& 0& 1& 1& 30& 2\\
N & 113& 0.013& 65& 1& 23& 4& 12& 1& 0& 2& 0& 0& 0& 23& 2\\
O & 123& 82& 0.007& 2& 40& 46& 17& 3& 0& 6& 8& 1& 3& 118& 11\\
Na & 29& 21& 20& 0.010& 5& 0& 3& 0& 0& 0& 1& 0& 0& 12& 2\\
Mg & 9& 5& 3& 0& 0.030& 1& 5& 0& 0& 0& 0& 1& 1& 34& 0\\
Al & 16& 11& 7& 0& 10& 0.025& 7& 0& 0& 1& 0& 0& 0& 14& 1\\
Si & 12& 7& 6& 0& 12& 1& 0.023& 0& 0& 1& 2& 0& 0& 17& 1\\
S & 19& 13& 21& 0& 13& 1& 4& 0.015& 0& 1& 0& 0& 0& 14& 1\\
K & 58& 45& 36& 1& 12& 1& 6& 0& 0.026& 1& 0& 0& 2& 12& 1\\
Ca & 13& 6& 1& 0& 3& 0& 4& 0& 0& 0.027& 1& 0& 0& 34& 0\\
Ti & 7& 9& 12& 0& 3& 0& 2& 0& 0& 0& 0.015& 0& 0& 12& 1\\
V & 36& 27& 23& 2& 4& 1& 11& 0& 0& 3& 1& 0.011& 2& 30& 1\\
Mn & 3& 4& 23& 0& 5& 1& 3& 0& 0& 0& 0& 0& 0.029& 5& 2\\
Fe & 12& 9& 8& 0& 8& 1& 7& 1& 0& 1& 6& 0& 1& 0.027& 2\\
Ni & 22& 7& 8& 0& 10& 1& 6& 1& 0& 1& 0& 0& 0& 46& 0.019\\

  \enddata
  \tablecomments{This table gives the changes in the spectrum of a
    $\teff = 4750\,\mathrm{K}$, $\logg = 2.5$ star with solar
    abundances due to abundance changes of $\pm0.1\dex$ for the 15
    elements considered in this paper. Each row weights the spectral
    changes due to each individual element using the weights for this
    row's element that I derived from the standard APOGEE weights in
    \appendixname~\ref{sec:windows}. Thus, the first line uses the
    weights for C to compute the weighted spectral variations due to
    different elements. The diagonal gives the actual weighted
    variation in the continuum-normalized spectrum, while the
    off-diagonal entries list the changes as a percentage of the
    diagonal entry. For example, N induces changes that are $34\,\%$
    of those induced by the same change in C when weighting using C
    weights. Using the set of weights from
    \appendixname~\ref{sec:windows}, the weighted spectral regions for
    all elements except for C, N, and O have only minor contributions
    from other elements.}\label{table:elemvar}
\end{deluxetable*}

APOGEE's abundance pipeline uses a set of numbers as a function of
wavelength for each element to weight the contribution of different
pixels to the $\Delta \chi^2$ when fitting for the abundance of that
element. These numbers give high weight to pixels that are highly
sensitive to the abundance of the element in question and not that
sensitive to the abundance of other elements. They are computed from
the derivatives of model spectra at $\teff = 4000\,\mathrm{K}$,
$\logg = 1$, and overall metallicity of $-2.0$, $-1.0$, and $0.0$ and
they also take into account how well a model for the spectrum of
Arcturus reproduces the high resolution, high signal-to-noise ratio
observed spectrum of Arcturus of \citet{Hinkle95a}, how well the whole
APOGEE sample is fit at each pixel, and how well the stars analyzed in
detail by \citet{Smith13a} are modeled. Full details on this procedure
are given in \citet{GarciaPerez15a}.

In this paper, I make use of these weights (also referred to as
``windows'') to analyze the spectral variations induced by abundance
changes of different elements. That is, when determining, for example,
the impact of Al variations, the APOGEE weights are used to only
consider wavelengths that are sensitive to the Al abundance. This
appendix describes some further analysis of the sensitivity of the
standard APOGEE DR12 windows to changes in the abundances of different
elements. The main purposes of this analysis are to specifically focus
on abundance changes near solar metallicity and to create a subset of
the standard windows that is less sensitive to the abundances of other
elements, in particular C and N.

\begin{figure*}[t!]
\begin{center}
\includegraphics[width=0.98\textwidth,clip=]{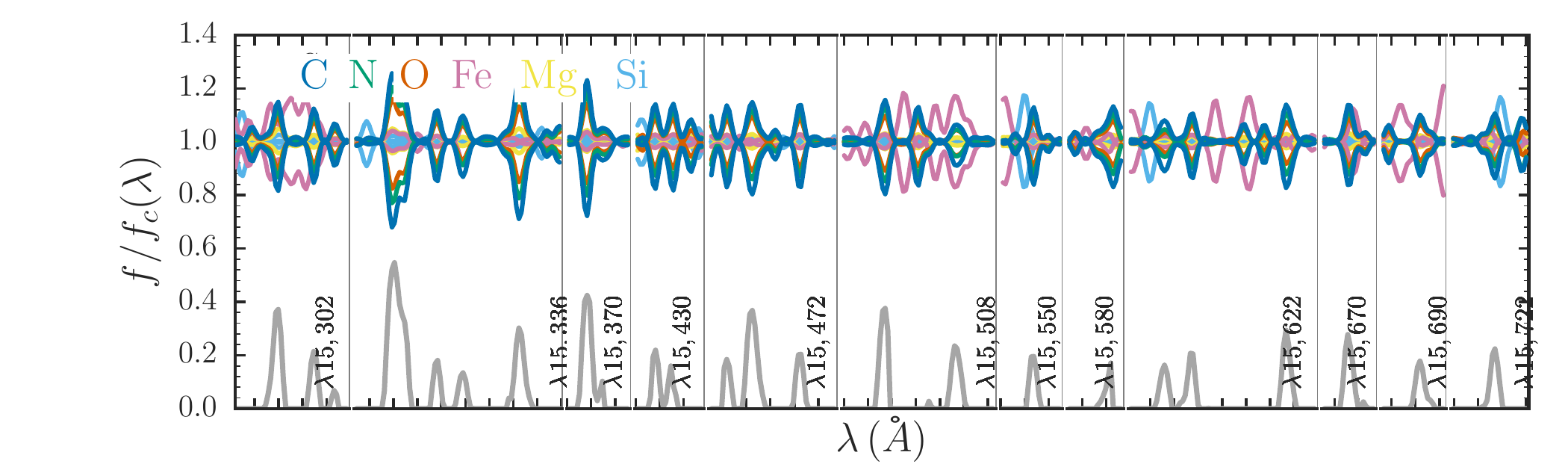}\\
\includegraphics[width=0.48\textwidth,clip=]{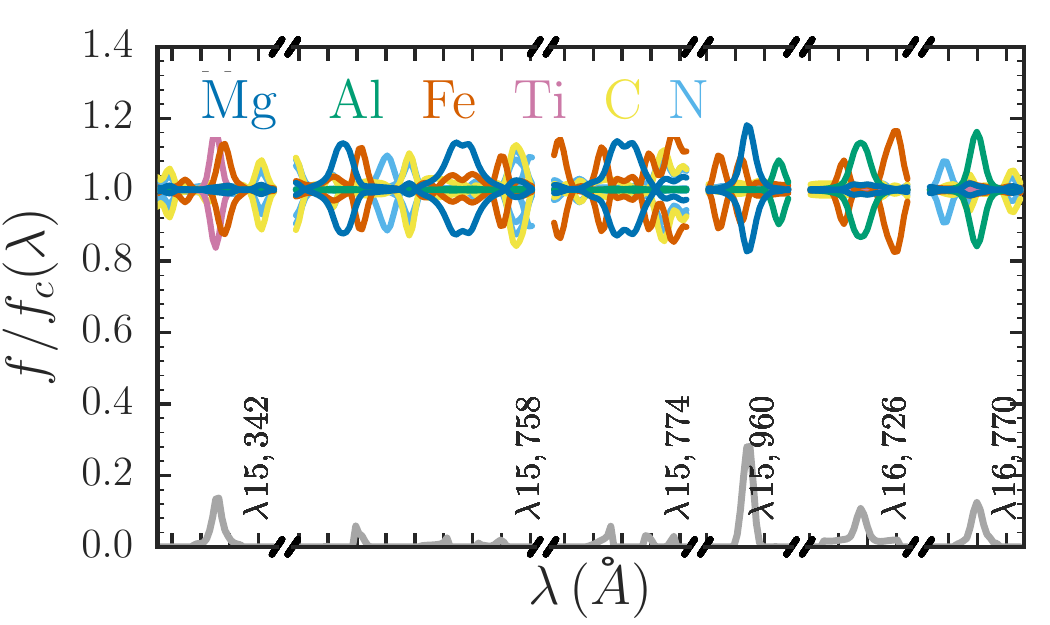}
\includegraphics[width=0.48\textwidth,clip=]{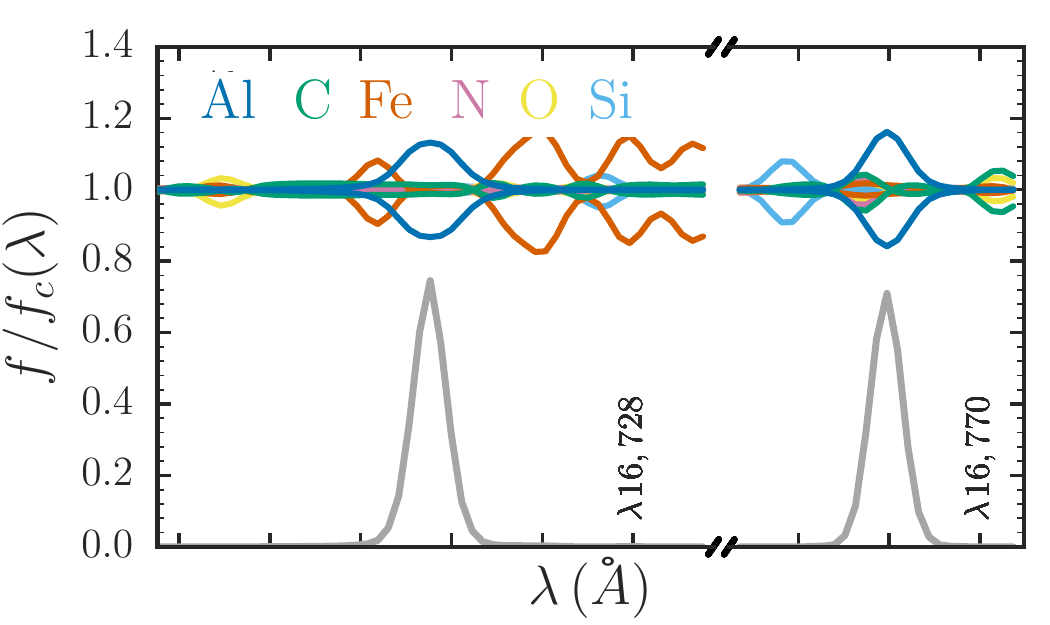}
\end{center}
\caption{Variations in the spectrum of a $\teff = 4750\,\mathrm{K}$,
  $\logg = 2.5$ star with solar abundances induced by $\pm0.1\dex$
  changes in the abundance of different elements. As examples, this
  figure shows those spectral variations in regions of the spectrum
  with C (top panel), Mg (bottom left panel), and Al (bottom right
  panel) absorption features; all variations are exaggerated by a
  factor of ten. The whole $H$-band wavelength range is broken into
  small regions and the separation in$x$ tickmarks is always $2\,\AA$;
  the wavelength of a single tickmark in each region is
  indicated. Variations due to elements other than C in the top panels
  and other than Mg or Al in the bottom panels are ordered by their
  weighted root-mean-square variation, computed using the standard
  APOGEE weights for C, Mg, and Al, respectively. These weights are
  displayed as the gray lines. Only half of the features for C are
  displayed here. C has ubiquitous absorption features, but
  disentangling them from those of N and O is difficult. Some of the
  Mg APOGEE weights cover wavelength regions where Mg does not have
  absorption features (such as the two reddest regions) and I remove
  these from consideration here.}\label{fig:elemvar}
\end{figure*}

To do this, I compute a baseline synthetic APOGEE spectrum using the
procedure described in \appendixname~\ref{sec:synspec} for a star with
$\teff = 4750\,\mathrm{K}$, $\logg = 2.5$, solar abundances,
microturbulence of $2\kms$, and convolving with the average LSF of all
APOGEE fibers. I then compute a set of spectra that vary the
abundances of all 15 elements considered in this paper separately by
$\pm0.1\dex$. The variations around the baseline spectrum are
displayed for a few example elements in
\figurename~\ref{fig:elemvar}. I then compute the root-mean-square
deviation for this $\pm0.1\dex$ change in all elements for each
element's windows, weighting by the APOGEE weights. For example, for
Al I use the Al weights and then compute the root-mean-square
variation in the spectrum for all 15 elements. This then returns the
effect of each individual element's abundance changes on the spectrum
near the Al absorption features. In \figurename~\ref{fig:elemvar}, the
element whose windows we are interested in is always displayed in blue
and other elements are ranked by the relative contribution to the
spectral scatter near the absorption features of that element.

\figurename~\ref{fig:elemvar} demonstrates that the standard APOGEE
weights for a given element (\eg, Mg) include many wavelength ranges
with significant contributions from other elements (\eg, the two
reddest Mg windows). To clean the list of windows for each element, I
compute the variation induced by other elements for each individual
window for that element (roughly, an individual absorption feature,
but for this can be quite extended for the molecular features). For
example, for each of the two Al windows in
\figurename~\ref{fig:elemvar}, I compute the variation induced by Al
and all other elements. I then remove any window that produces less
than a 0.01 change in the continuum-normalized spectrum when varying
the abundance of that window's element by $\pm0.1\dex$ or if any of
the other elements induce a variation greater than 34\,\% of that
induced by the window's corresponding element. For example, if the
variation due to Al in the first Al window is 0.005, the window would
be removed (this is not the case). Or if another element, say Mg,
creates a spectral variation larger than 34\,\% of that of Al in this
window, the window would be removed (also not the case). Many of the
elements require some fine tuning of these cuts to not remove too many
individual windows: K is kept, because there is only a single line in
the spectrum, other elements are allowed to contribute up to 100\,\%
or 200\,\% for C and N, respectively (basically, because of the
ubiquitous CN features), O is kept down to changes as small as 0.005
and up to contributions of other elements of 500\,\% (because most O
features are weak). Additionally, for Na, Ti, V, Mn, and Ni I only
consider the contribution of other elements and use a cut-off of
34\,\%, 30\,\%, 40\,\%, 25\,\%, and 50\,\%, respectively. While these
cuts are somewhat arbitrary, they have been chosen to keep a
reasonable number of windows for each element that are not too
affected by variations in other elements.

The weighted variations induced in the spectrum by $\pm0.1\dex$
abundance changes for the final set of weights are given in
\tablename~\ref{table:elemvar}. Along the diagonal, this table gives
the magnitude of spectral changes for each element weighted by the
weights for that element. The off-diagonal entries show how much
variation abundances changes in other elements induce, given as a
percentage of the main element's variation. This demonstrates that we
end up with a relatively clean set of weights for each element. Most
$\pm0.1\dex$ abundance changes induce weighted spectral variations of
about 0.01 to 0.025, which is larger than the typical noise of each
pixel. This weighted spectral variation, however, does not show the
number of pixels at which such large variations exist. Elements with a
large number of pixels with non-zero weights will lead to stronger
constraints on abundance variations.

\end{document}